\documentclass[journal]{IEEEtran}

\ifCLASSINFOpdf
\else
\fi

\usepackage{lipsum} 

\usepackage{tikz,xcolor,hyperref}

\definecolor{lime}{HTML}{A6CE39}
\DeclareRobustCommand{\orcidicon}{%
	\begin{tikzpicture}
	\draw[lime, fill=lime] (0,0) 
	circle [radius=0.16] 
	node[white] {{\fontfamily{qag}\selectfont \tiny ID}};
	\draw[white, fill=white] (-0.0625,0.095) 
	circle [radius=0.007];
	\end{tikzpicture}
	\hspace{-2mm}
}

\foreach \x in {A, ..., Z}{%
	\expandafter\xdef\csname orcid\x\endcsname{\noexpand\href{https://orcid.org/\csname orcidauthor\x\endcsname}{\noexpand\orcidicon}}
}

\usepackage{epsfig,makeidx,color}
\usepackage{amsmath,amssymb,bbm}
\usepackage{cite,graphicx,lipsum}
\usepackage{tabularx} 
\usepackage{float}    
\usepackage{rotating}
\usepackage{enumerate}
\usepackage{booktabs}
\usepackage{adjustbox}
\usepackage{soul}
\usepackage[switch,pagewise]{lineno}
\usepackage[caption=false]{subfig}
\usepackage{multirow}

\title{A Survey on IoT Ground Sensing Systems for Early Wildfire Detection: Technologies, Challenges and Opportunities}
\author{Chiu Chun {Chan}\orcidA{}, \textit{Member, IEEE}, Sheeraz {A. Alvi}\orcidB{}, \textit{Member, IEEE}, Xiangyun {Zhou}\orcidC{}, \textit{Fellow, IEEE},\\ Salman {Durrani}\orcidD{}, \textit{Senior Member, IEEE}, Nicholas {Wilson}\orcidE{}, Marta {Yebra}\orcidF{}\\

\thanks{
\tiny{This research was funded by the Bushfire Research Centre of Excellence, which was co-sponsored by the Australian National University and Optus. The Australian Capital Territory Emergency Services Agency and the Australian Capital Territory Parks and Conservation Service provided support for the collection and organisation of ignition data. We thank Sam Beaver for organising the experimental burns and Jyotirao Wankhede for assisting with the experimental burns. We also thank Dr. Andrew Sullivan and the CSIRO Bushfire Behaviour and Risks team for their help in initiating the testing of the sensing modules in the Pyrotron Lab. (Corresponding author: Chiu Chun Chan)

Chiu Chun Chan, Sheeraz A. Alvi, Xiangyun Zhou, Salman Durrani and Marta Yebra are with the School of Engineering, The Australian National University, Canberra, ACT 2601, Australia (e-mail:
\{chiuchun.chan; sheeraz.alvi; xiangyun.zhou; salman.durrani\}@anu.edu.au).

Nicholas Wilson and Marta Yebra are with Fenner School of Environment and Society, The Australian National University,  Canberra, ACT 2601, Australia (e-mail:\{nicholas.wilson; marta.yebra\}@anu.edu.au).

All authors are with the Bushfire Research Centre of Excellence, The Australian National University, Canberra, ACT 2601, Australia.

©2024 IEEE. Personal use of this material is permitted. Permission from IEEE must be obtained for all other uses, in any current or future media, including reprinting/republishing this material for advertising or promotional purposes, creating new collective works, for resale or redistribution to servers or lists,or reuse of any copyrighted component of this work in other works.
}

} 
}

\begin{document}
\maketitle
\begin{abstract}
The threat posed by wildfires or bushfires has become a severe global issue due to the increase in human activities in forested areas and the impact of climate change. Consequently, there is a surge in the development of automatic wildfire detection methods. Approaches based on long-distance imagery from satellites or watchtowers encounter limitations, such as restricted visibility, which results in delayed response times. To address and overcome these challenges, research interest has grown in the implementation of ground-based Internet of Things (IoT) sensing systems for early wildfire detection. However, research on energy consumption, detection latency, and detection accuracy of IoT sensing systems, as well as the performance of various anomaly detection algorithms when evaluated using these metrics, is lacking. Therefore, in this article, we present an overview of current IoT ground sensing systems for early wildfire detection. Camera and environmental sensing technologies suitable for early wildfire detection are discussed, as well as vision-based detection algorithms and detection algorithms for environmental sensing. Challenges related to the development and implementation of IoT ground sensing systems for early wildfire detection and the future research directions important for creating a robust detection system to combat the growing threat of wildfires worldwide are discussed.
\end{abstract}
\begin{IEEEkeywords}
IoT, sensing, wildfire, bushfire, forest fire, early fire detection.
\end{IEEEkeywords}
\section{Introduction}\label{Sec_Intro}
\IEEEPARstart{W}{ildfire}, also widely known as bushfire in Australia and forest fire in Europe, present a significant risk to people, the environment and assets. As of 2023, there has been an increase of $1.5^\circ$C in the global surface air temperature compared to the 1850-1900 baseline, attributed to climate change, leading to more days with high fire risk, additionally, the ongoing El Ni\(\mathrm{\Tilde{n}}\)o event has exacerbated conditions, elevating temperatures to 1.6\(^\circ\)C above pre-industrial levels by early 2024~\cite{StateoftheClimate2024}. As fire weather worsens, fire occurrence increases~\cite{CLARKE201934,fire6050195} and the window of opportunity for successful containment and suppression after ignition narrows~\cite{RODRIGUES2019915,10.5849/forsci.10-096}. Hence, rapid detection of new ignitions will become an increasingly important strategy for fire managers as the climate warms.
\vspace{-1em}
\subsection{Impact of Wildfire Across the Globe}
During the 2019-2020 period in Australia, widespread fires known as the ``Black Summer'' affected various states, burning approximately 24 million hectares, resulting in 33 deaths and around 450 injuries due to smoke inhalation~\cite{_2021_australias}. Besides, the impact on wildlife was immense. According to the World Wide Fund for Nature, billions of animals were injured, killed or displaced~\cite{vernick_2020_3} due to the Black Summer. For the already fragile ecosystem of animals in Australia, this is a major hit, as many animals, including koalas, wombats and wallabies, are listed as endangered or vulnerable~\cite{_2019_threatened}.

Worldwide, wildfires are intensifying. According to the European Forest Fire Information System, in 2023 alone, wildfires burned an estimated 908,368 acres of land in countries of the European Union, with regions such as Greece experiencing severe damage and around 20 million tonnes of carbon dioxide emitted~\cite{centerfordisasterphilanthropy_2023}. In June 2023, major wildfires erupted in the northeastern areas of Kazakhstan, resulting in the highest annual death toll. At least 15 people lost their lives while fighting the fires~\cite{kazakhstanforestfire}. In 2020, the United States experienced the most devastating wildfires in recent history, with more than 50,000 fires burning approximately 3.64 million acres of land, which is the highest figure recorded in the 24-year period from 2000 to 2024.~\cite{centerfordisasterphilanthropy_2020_2020}. Despite the consistent number of wildfires in the past three decades, the annual burnt area has increased by 1.62 million hectares from the levels seen in the 1990s in the United States~\cite{wildfirestat}. Similarly, Canada experienced a significant number of wildfires, with around 5,500 fires destroying 17.34 hectares of land~\cite{canadafirestat}. The Brazilian National Institute of Space Research has revealed 3.36 million hectares of land burnt in Brazil annually since 2003~\cite{ye2021risk}. In addition to these observed increases in fire activity, predicted warming and drying of the climate across much of the world are likely to increase fire activity and result in long-term declines in forest carbon storage, further warming the climate~\cite{clarke2022forest}. In light of the increasing threat posed by wildfires, the imperative to develop precise and reliable detection technologies has taken precedence. 

\begin{table*} 
    \centering \vspace{-1em}
        \caption{Example of existing survey papers on technological solutions for wildfire detection and the contribution of this article}
        \resizebox{\textwidth}{!}{
    \begin{tabular}{l l c c c c c }
    \multirow{2}{*}{Existing Works} & \multirow{2}{*}{Summary} & \multicolumn{4}{c}{Platforms} \\
    & & & Satellite Remote Sensing & UAV & Camera Watchtower & IoT\\
    \toprule
    Panagiotis \textit{et. al.}~\cite{barmpoutis2020review} & A review and comparison of various imaging-based technologies for early detection of wildfires & & \checkmark & \checkmark & \checkmark & \checkmark\\
    
    \multirow{2}{*}{Ankita \textit{et. al.}~\cite{su141912270}} &  A review of different technologies for early wildfire detection & & \multirow{2}{*}{\checkmark} & \multirow{2}{*}{\checkmark} & \multirow{2}{*}{\checkmark} & \multirow{2}{*}{\checkmark} \\ 
    & and the comparison of various commercially available early wildfire detection systems & & & & \\
    
    \multirow{2}{*}{Rafik \textit{et. al.}~\cite{10.1007/978-3-030-21005-2_32}} & An overview of different imaging-based technologies used for the early detection of wildfires && \multirow{2}{*}{\checkmark} & \multirow{2}{*}{\checkmark} & \multirow{2}{*}{\checkmark} & \multirow{2}{*}{}\\
    & and the application of DL algorithms in fire imaging detection & & & & \\
    
    Ahmad \textit{et. al.}~\cite{doi:10.1155/2014/597368} & A review and comparison of various technologies used for early wildfire detection & & \checkmark  & & \checkmark & \checkmark\\
    
    Faroudja \textit{et. al.}~\cite{abid2021survey} & A review and comparison focusing on ML/DL algorithms suitable for early detection of wildfires & & & & &\\
    
    Francesco \textit{et. al.}~\cite{s23146635} & A review of ML/DL algorithms suitable for imaging-based early detection of wildfires & & & \checkmark &  \checkmark& \\
    
    Abdelmalek \textit{et. al.}~\cite{BOUGUETTAYA2022108309} & Comparison of ML/DL algorithms suitable for early wildfire detection on UAVs & & & \checkmark\\
    Hong \textit{et. al.}~\cite{hong2023wildfire} & Comparison of ML/DL algorithms suitable for early wildfire detection on camera watchtowers & & & & \checkmark \\
    
    \multirow{2}{*}{Mounir \textit{et. al.}~\cite{grari2022using}} & A review on early wildfire detection using IoT systems  & & & & & \multirow{2}{*}{\checkmark} \\
     & and discussion on challenges in IoT wildfire detection & & & & \\
    
    \multirow{2}{*}{This article} & An overview of IoT sensing technologies suitable for early wildfire detection &&&&& \multirow{2}{*}{\checkmark}\\
        & and the analysis of the challenges in implementing such a system & & & & \\
    
    \bottomrule
    \end{tabular}}
    \label{tab:literature review} \vspace{-1em}
\end{table*}

\begin{table}
    \centering
        \caption{Table of acronym}
        \begin{tabular}{l l}
    \toprule
    Artificial Neural Network & ANN \\
    The Australian National University & ANU \\
    Autoregressive Integrated Moving Average & ARIMA\\
    Backpropagation & BP \\
    Convolution Neural Network & CNN\\
    Carrier-Sense Multiple Access with Collision Avoidance & CSMA/CA \\
    Cumulative Sum Control Chart & CUSUM\\
    Deep Learning & DL\\
    Decision Tree & DT\\
    Discrete Wavelet Transform & DWT \\
    End Node & ED\\
    Gas-Sensitive Fleid Effect Transistor & GasFET \\
    Gaussian Mixture Model & GMM \\
    Histogram of Optical Flow & HOF \\
    Histogram of Orientated Gradient & HOG \\
    Internet of Thing & IoT\\
    Infrared & IR \\
    K-Nearest Neighbours & KNN \\
    Local Outlier Factor & LOF \\
    Line-of-Sight & LoS \\
    Low Power Wide Area Network  & LPWAN \\
    Long-Short-Term Memory Network & LSTM \\
    Medium Access Control & MAC \\
    Multi-Criteria Decision Analysis & MCDA \\
    Microcontroller & MCU \\
    Machine Learning & ML \\
    Maximum Power Point Tracking & MPPT\\
    Metal Oxide Semiconductor & MoS \\
    Non-Dispersive Infrared & NDIR\\
    Passive Infrared & PIR \\ 
    Particulate Matter & PM \\
    Radial Basis Function & RBF \\
    Random Forest & RF \\
    Relative Humidity & RH \\
    Recurrent Neural Network & RNN \\
    Resistive Temperature Detector & RTD \\
    Seasonal Autoregressive Integrated Moving Average & SARIMA \\
    Scale-Invariant Feature Transform & SIFT \\
    Support Vector Machine & SVM \\
    Total Volatile Organic Compounds & TVOC \\    
    Unmanned Aerial Vehicles & UAV\\
    Ultraviolet & UV \\
    \bottomrule
    \end{tabular}
    \label{tab:acronym} \vspace{-2em}
\end{table}
\vspace{-1em}
\subsection{Early Wildfire Detection: Existing approaches}
The ability to detect wildfires in their early stages allows effective and efficient fire suppression~\cite{informit}. Specifically, early detection of initial fire stages with benign fire behaviour, enables the timely deployment of fire suppression tactics to increase the probability of initial attack success~\cite{10.5849/forsci.10-096}. Therefore, it is vital to develop early wildfire detection methods and technologies. Here, we present a set of requirements to gauge the capability of existing technologies for early wildfire detection. The algorithm should be able to identify fires while they are still small enough to be successfully contained or suppressed. As the fire grows larger, it may become too difficult to extinguish~\cite{10.5849/forsci.10-096}. Therefore, early detection that can reduce the time it takes to detect and locate small fires is hugely beneficial. Moreover, wildfire detection systems based on environmental monitoring are highly desirable due to the correlation between fire hazards and changes in the environment. Finally, the scalability of wildfire detection systems is beneficial so that systems can not only be deployed on a large scale but also be deployed in different terrains. 

We hereby briefly introduce the existing technologies for wildfire detection and analyse their suitability for early wildfire detection. For nearly a century, \textbf{watchtowers} and public reporting have been used to detect wildfires. Unfortunately, watchtowers require observers to remain on top of the tower during certain hours of the day and specific periods of the year, constantly looking for smoke that implies fire and the effectiveness of public reporting is significantly influenced by the population density of the region. With the advancement of computer vision technologies~\cite{yan2022transmission,doi:10.1126/sciadv.abn9328,doi:10.1126/scirobotics.abl7755}, fire detection from watchtowers can now be accomplished day and night with the help of arrays of high-resolution vision / thermal cameras that record visual information, which is then processed using different detection algorithms to recognise smoke, flames, or heat from fire. However, there are some challenges associated with this technology for early wildfire detection. Firstly, fires take some time before producing a signal that is strong enough to detect from a distance, which can prolong their detection time. Secondly, monitoring hilly terrain becomes difficult due to line-of-sight blockage. Consequently, it may be difficult to detect small fires in these areas. Thirdly, this technology lacks the ability to convey environmental changes, such as changes in relative humidity and temperature, which cause diurnal changes in fuel moisture and associated fire activity~\cite{nolan2016large}.

As another detection technology, the \textbf{deployment of crewed fire-spotting aircraft} over lightning strike areas or areas of high fire risk is also used, but is often delayed due to resourcing issues or restrictions on night-flying operations. The use of long-endurance Unmanned aerial vehicles (UAVs) equipped with thermal cameras is on the rise, mainly due to their availability to operate in difficult weather conditions, day and night, helping to overcome some of the limitations associated with traditional fire-spotting aircraft~\cite{9424181,doi:10.1139/juvs-2020-0009,afacreport}. UAVs offer the potential to rapidly confirm suspected ignitions and provide information on the size and behaviour of fires, particularly fires in remote locations. However, authorisation is required and regulations must be followed to operate this system to detect potential fires.

For many years, \textbf{satellite imagery} has been a valuable tool for monitoring large-scale wildfires mainly based on thermal anomalies detected by thermal sensors~\cite{chuvieco2020satellite}. But it also comes with limitations, particularly in terms of temporal or spatial resolution of such sensor systems, making it challenging to detect fires when they are still small. The Low Earth Orbit satellite constellation network offers relatively high spatial resolution capable of detecting small fires~\cite{SCHROEDER2016210}. However, the temporal frequency is entirely dependent on the size of the constellation, and ensuring prompt detection can be expensive due to the costs associated with deploying and maintaining a large constellation. In addition, opting for a smaller constellation may compromise the reliability of fire detection~\cite{SMITH200795}. Novel algorithms developed specifically for early fire detection using geostationary satellites, which offer temporal resolution of minutes and spatial resolution of kilometres, may offer advances, but have not yet been thoroughly tested~\cite{doi:10.1080/15481603.2022.2143872}.

In addition to the technologies mentioned above, a promising complementary approach to detect early wildfires is to use \textbf{ground sensing systems based on the Internet of Things (IoT)}. These systems provide real-time data from a variety of ground sensors at desired densities for timely and accurate fire detection, due to the proximity of the ground sensors to the fire ignition locations. This IoT-based system can autonomously raise early alarms for wildfire detection, identify its location, and provide situational awareness by reporting the fire movement. The communication delay from data generation at the ground sensors to reception at the central node can be in a few minutes. Moreover, this kind of system can complement other fire detection and extinguishing systems by identifying the area of interest, which then can be investigated for further details and validation of ignitions as well as to initialise responses, or to monitor areas that are difficult to monitor with other technologies. The benefits of a ground sensing systems would also go beyond wildfire detection and suppression, as they can also be used for environmental monitoring purposes such as fuel moisture monitoring. Due to these unique advantages,  many IoT companies have integrated IoT-based wildfire detection systems into their offerings~\cite{Dryad,Milesight,attentis, n5sensor}. For example, N5 Sensors offers a system that uses their cutomised gas sensors to identify various volatile compounds, carbon monoxide, and CO$_2$, with the commitment to detect fire ignitions in five minutes~\cite{n5sensor}. Another example is presented by Dryad, where the Bosch BME680 sensor is used in its IoT setup to detect wildfires, ensuring detection times of under an hour~\cite{Dryad}. However, it is essential not to rely solely on one detection system for early fire detection, but to recognise the benefits of combining different wildfire detection approaches and take a more comprehensive approach. This requires long-term research and development efforts across academia and industry on a large scale. In Australia, a leading effort on this front is the recent establishment of the Bushfire Research Centre of Excellence supported by the Australian National University~(ANU) and Singtel Optus Pty. Ltd. The centre aims to develop an integrated and layered solution to detect and extinguish small fires using multiple technologies, some mentioned above~\cite{informit,afacreport}.

\subsection{Prior Work and Our Contributions}

Wildfire detection has been an active topic of research over the past two decades, and several surveys have been published on this general topic. These are summarised in Table~\ref{tab:literature review}, which highlights the objectives of each survey paper and the specific technological areas on which they focus. Many survey papers, such as \cite{barmpoutis2020review, su141912270} and other survey papers cited within these works, provide information on various technologies applicable to early wildfire detection. These surveys evaluate the strengths and weaknesses of different technologies, including the integration of Machine Learning (ML) and Deep Learning (DL) methods for early wildfire detection and prevention. Other survey papers, such as~\cite{barmpoutis2020review,BOUGUETTAYA2022108309,s23146635} and~\cite{hong2023wildfire}, cover satellite-, UAV- and camera-watchtower-based detection techniques, and few survey papers do cover ground-based IoT systems for early wildfire detection~\cite{su141912270,doi:10.1155/2014/597368,10.1007/978-3-030-21005-2_32,grari2022using,s23146635,abid2021survey}. In~\cite{su141912270,doi:10.1155/2014/597368,10.1007/978-3-030-21005-2_32}, summaries of recent advances in early fire detection and prevention technologies are provided, with minimal coverage of IoT systems. In~\cite{grari2022using}, reviews of IoT-based wildfire detection using ML and DL approaches are discussed, together with challenges in IoT wildfire detection. However, in~\cite{su141912270,doi:10.1155/2014/597368,10.1007/978-3-030-21005-2_32,s23146635,abid2021survey} and~\cite{grari2022using}, the challenges faced in the deployment of IoT sensing systems are only briefly mentioned. This is an important gap in the literature that this article aims to address. 

In this article, our objective is to explore the potential benefits of modern sensing technologies and their compatibility with IoT sensing systems to detect early-stage wildfires. Specifically, we investigate the use of vision-based and environmental monitoring technologies, combined with anomaly detection algorithms, to detect wildfires in their early stages. Furthermore, the study highlights critical factors and challenges to consider when implementing IoT-based sensing systems for reliable and efficient early wildfire detection, such as sensor placement and network connectivity. The objective of this article is to provide a comprehensive understanding and guidance for the implementation of IoT ground sensing systems for early wildfire detection, together with its challenges and possible research extensions. 

\begin{figure*}
    \centering
    \includegraphics[width=\textwidth]{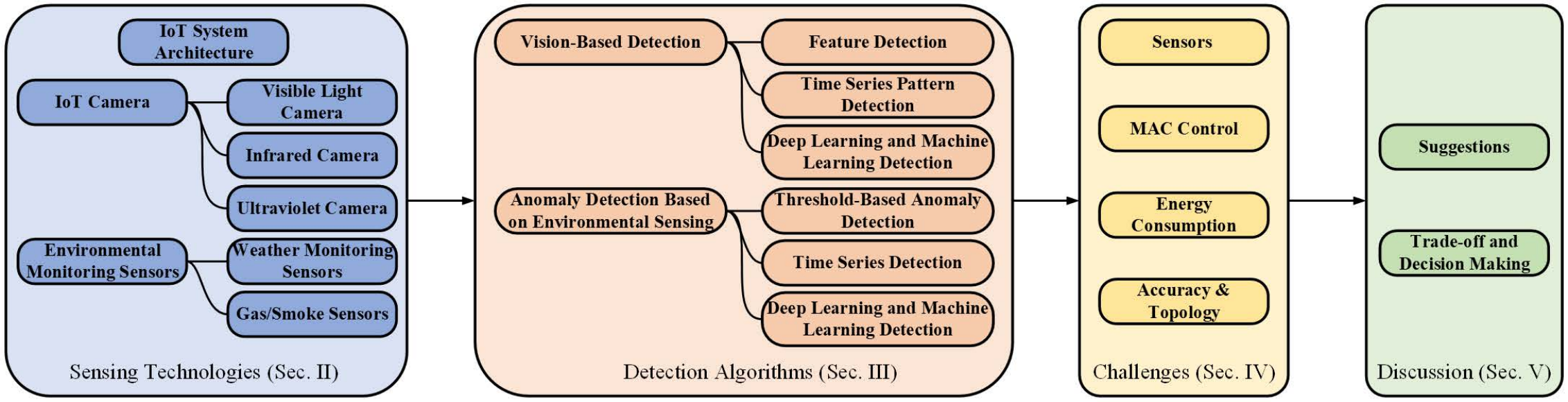}
    \caption{Structure of this article about IoT ground sensing systems for early wildfire detection.}
    \label{Fig:structure} \vspace{-1em}
\end{figure*}

The contributions of this work are summarised as follows:
\begin{itemize}
    \item In-depth technical overview and performance analysis of existing vision-based and environmental sensing technologies using experimental studies and summarising from the literature.
    \item Identification of appropriate sensors for early wildfire detection under a resource-constrained requirement in terms of detection time, energy consumption, and cost.
    \item An overview of existing detection techniques using vision and environmental sensing data. 
    \item Identification of operational challenges associated with the deployment of IoT systems in remote areas.
    \item Recommendations for developing IoT ground sensing systems in early wildfire detection, along with suggestions to help decision makers identify appropriate deployment strategies according to specified performance metrics.
\end{itemize}

\subsection{Paper Organisation}
The structure of this article is outlined in Figure~\ref{Fig:structure}. Section~\ref{Sec_lit_review_new} discusses the architecture of IoT systems and IoT camera modules and environmental monitoring sensors that can be used for such an application. It also provides a brief overview of the operating principles of different types of sensors, their characteristics and use case examples. Finally, the performance of the sensors is compared in multiple aspects. In Section~\ref{Sec_Algor}, we provide an overview of the wildfire detection algorithms commonly used in IoT systems. These algorithms can be divided into two categories: vision-based detection and anomaly detection based on environmental sensing data. We further classify these detection methods based on their complexity, such as feature detection, time series pattern detection and ML \& DL detection in vision-based and threshold detection, time series detection, ML \& DL detection in environmental sensing. Additionally, we discuss energy consumption, detection delay, requirements for deploying detection algorithms and the popularity of different detection algorithms is also provided at the end of the section. Section~\ref{Sec_Challenges} explains the operational challenges associated with the deployment of IoT applications for wildfire detection, including sensor limitations, energy efficiency considerations, network coverage constraints, the accuracy of detection algorithms and the challenges related to system deployment. Finally, Section~\ref{Sec_pos_sol} proposes suggestions for the design and deployment of IoT ground sensing systems in the context of early wildfire detection, offering solutions to most of the challenges identified in Section~\ref{Sec_Challenges}. Recognising the inherent trade-offs among some of these challenges, this section also introduces a performance metric that can guide IoT designers in their decision-making processes. Furthermore, it explores potential extensions for future research within this domain. The list of common acronyms used in this article is presented in Table~\ref{tab:acronym}.

\section{IoT Technologies for Wildfire Sensing}\label{Sec_lit_review_new}
\begin{figure}
    \centering
    \includegraphics[width=\linewidth]{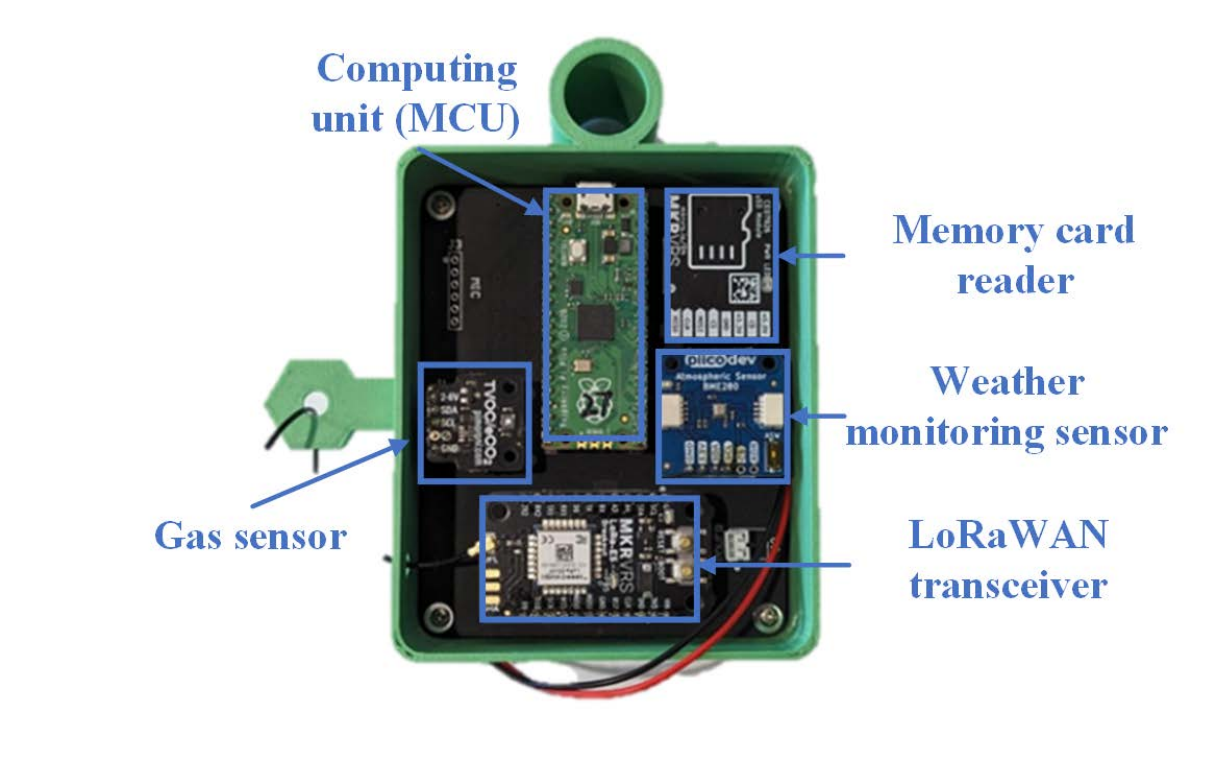}
    \caption{ A hardware prototype of an IoT end device develpoed by the Bushfire Research Centre of Excellence, which includes a MCU, an memory card reader, a gas sensor, a weather monitoring sensor and a LoRaWAN transceiver.}\vspace{-1em}
    \label{Fig:hardwareprototype}
\end{figure}

IoT systems are typically composed of three core elements: (i) end nodes (EDs), (ii) gateways (also referred to as aggregators) and (iii) the central node (also referred to as cloud server). The EDs consist of ground sensors, including low-cost camera modules and / or environmental monitoring sensors, interconnected with microcontrollers (MCUs). These MCUs typically perform data pre-processing tasks using simple algorithms such as moving averages, threshold detection, linear regression, or ML classification. The primary purpose of data pre-processing is to improve data quality, consistency and analysis readiness before sending it to the gateway. After completion of the data processing phase, the data are transmitted to the gateway through suitable Low Power Wide Area Network (LPWAN) communication technologies customised to specific user requirements. This transmission can be optimised based on factors such as data rates, extended ranges, or energy efficiency. An example of a hardware prototype of the IoT ED for early wildfire detection is illustrated in Fig.~\ref{Fig:hardwareprototype}. 

The gateway then forwards the vital data received to the central node for post-processing. During the post-processing stage, advanced technologies with high computing capabilities, such as DL algorithms, can be employed at the central node. Unlike EDs, the central node is not constrained by energy consumption or computational capacity.  These advanced technologies enable a comprehensive analysis of sensor data, facilitating decision-making processes, such as the detection of fire alarms in a specific scenario~\cite{hanes2017iot}. However, the level of accuracy and energy efficiency required for such a wildfire detection system remains largely an open and active research issue. Developing a robust IoT sensing system for early fire detection requires careful consideration of these performance characteristics and metrics such as overall energy consumption, response time and accuracy detection can be used to evaluate its efficiency.

In this section, we provide a detailed discussion of sensor technologies crucial for early wildfire detection, focusing on the principles, features and applications of different IoT sensors. The comparison of commercially available sensor modules are provided, highlighting their use in prior wildfire detection projects. We categorise ground sensors used in wildfire detection into two main categories: vision sensors and environmental monitoring sensors. These sensors collect a variety of data types, including visual recordings and meteorological information of forested areas, for local analysis on EDs or for centralised processing. This section also reviews studies on the integration of these sensors with IoT systems, evaluating their effectiveness and practicality for early wildfire detection.

\subsection{Vision-Based Ground Sensing Technologies}

Vision-based fire detection has been a longstanding solution in IoT application, employed across both indoor and outdoor settings. Its interpretability by both humans and machines has solidified its status as a reliable detection method. We categorise the vision-based sensors for wildfire detection into three categories spaced across the electromagnetic spectrum.

\subsubsection{\textbf{Visible Light Range Cameras}}

Numerous research studies have integrated visible range (RGB) camera systems into ground IoT sensing systems to enable fast and robust wildfire detection. Unlike long-range and high-resolution cameras used on watchtowers or satellites, IoT ground sensing systems require small, cost-effective and energy-efficient cameras for scalable deployment. Recent advances in semiconductor technologies and compact lens systems have made it feasible to miniaturise camera modules. To make it more feasible for IoT applications, these low-performance camera modules have also been modified to operate at smaller viewing angles, lower resolutions and frame rates, to reduce energy consumption and the need for large lens systems. For example, these camera modules only provide video quality at a maximum of 1080p30fps with 102$^\circ$ viewing angle. Such camera modules have been used in various IoT projects, including those involving wildfire detection~\cite{KANAKARAJA20213907,8421338,10.1007/s10586-019-02981-7,8777414}. Notable examples of IoT cameras include IMX219~\cite{IMX219} and OV7670~\cite{OV7670}. Using these IoT cameras, microcontrollers can capture 2D video streams or images in pixel-matrix format, enabling vision-based fire detection. However, camera images are sensitive to adverse weather conditions and obstructions and are less reliable in dim conditions, particularly for early fire detection. Additionally, the limited viewing angle resulting from camera miniaturisation requires the use of a rotating platform or multiple cameras to achieve a 360$^\circ$ view if the detection system is based solely on cameras.

\subsubsection{\textbf{Infrared Range Cameras}}
To improve the night visibility of these ground-sensing devices, the implementation of infrared (IR) cameras has been explored as an alternative solution. This approach of paralleling the utilisation of IR technology with visible light cameras has shown a significant improvement in night visibility in watchtowers and has been adopted by many commercial products in fire detection~\cite{attentis,exci}. IR cameras capture thermal images, where each pixel directly represents the temperature or heat intensity of the corresponding area. This characteristic simplifies the fire detection process, providing an intuitive indication of the presence of a fire. In an extremely simplified scenario, fire detection can be achieved by analysing the histogram or sum of the pixel intensities. However, the adoption of IR cameras also comes with a drawback, as they tend to be more expensive compared to regular cameras due to their limited market demand. Although IR cameras facilitate fire detection at night, their reliability can be compromised if the detection distance and temperature range of IR sensors are insufficient. For example, certain IoT compatible IR cameras may only detect temperature ranges from 0-80 $^\circ$C, while forest soil can reach 70~$^\circ$C during high-hazard weather conditions. Therefore, relying solely on IR cameras in an IoT system may not be the optimal choice. In~\cite{goyal2020yolo}, authors proposed a fire detection system that can be applied to IoT or UAVs, and the experimental results show its feasibility by combining the IR camera and the DL detection algorithm. 

\subsubsection{\textbf{Ultraviolet Range Cameras}}
In addition to visible-light and IR cameras, ultraviolet (UV) cameras have also shown fire-detecting capability in situations where smoke or other factors may obscure the visible flames. In~\cite{6733740,ALHAZA20152343}, similar indoor firefighting robot designs have been explored, using low-cost MCUs and multispectral detection methods, including UV, IR and visible light. However, it is important to note that UV cameras are not the primary method of fire detection used in most fire safety systems. They are generally used together with other detection methods because the UV detection range and accuracy are lower due to the high UV absorption by particles and false positive detection caused by other UV-emitting or UV-reflective sources. Furthermore, the requirement for high-quality components, such as lenses, filters and photo sensors, in the construction of UV cameras makes them significantly more expensive than other types of cameras. To address this issue, some researchers have tried to reduce the cost of UV cameras by eliminating colour filters from IoT cameras to allow their UV capture features, and these cost-effective UV cameras have demonstrated their potential in chimney smoke detection~\cite{s16101649}.

\subsection{Environmental Monitoring Technologies}
As mentioned in the previous subsection, vision-based approaches based on visibility from the monitoring location may face potential detection delays. To address this, environmental monitoring sensors offer an alternative by providing information from different perspectives on fire hazards and incidents. These sensors offer advantages such as smaller size, lower cost, reduced energy consumption and increased autonomy compared to the vision-based technology. In addition, they do not require a direct line-of-sight (LoS) between EDs and the monitoring location, enabling faster detection based on changes in the environment. Therefore, environmental monitoring sensing technologies have gained prominence in wildfire detection projects. These projects commonly employ front-line sensors, including thermal, humidity, pressure, gas and smoke sensors.

\subsubsection{\textbf{Weather Monitoring Sensor}}
The utilisation of temperature, humidity and air pressure detection technologies is prevalent in IoT projects, with thermistors and resistive temperature detectors (RTDs) commonly used for temperature sensing~\cite{s19183905,8058280}. Similarly, humidity and air pressure sensing are employed for resistive humidity sensing, thermal conductivity humidity sensing and piezoresistive pressure sensing, respectively, considering factors such as size and cost in mass production. Weather sensors have the potential to be an important asset in monitoring fire risk, as temperature and relative humidity can be used to calculate the moisture content of fire fuels, which is a major factor in controlling fire activity~\cite{RESCODEDIOS201564}, including ignition probability~\cite{CAWSON2022120315}. Furthermore, it allows the detection of indicators such as increased temperature, humidity reduction and air pressure changes resulting from fire heat, which can be observed by EDs. These indicators are commonly used to determine the spread of wildfires, as well as widely used in various indoor fire detection schemes~\cite{8093388,s22093310}.

Thermistors exhibit high sensitivity to temperature changes as a result of the heat conductivity of their sensing materials, resulting in a lower negative temperature coefficient and a nonlinear response. However, thermistors have limited operating temperature ranges and may not be optimal as standalone temperature sensors for IoT fire detection. On the other hand, RTDs offer wider temperature operating ranges but have lower sensitivity compared to thermistors. In humidity detection, both capacitive and resistive humidity sensors are cost-effective. However, capacitive humidity sensors are sensitive to contaminants and temperature variations, while resistive humidity sensors are generally larger in size, which can impact the design considerations of IoT projects. \cite{8052688,9992560} have demonstrated the potential application of these sensing technologies to detect fire hazards in forested areas. A recent study~\cite{SensorComparsion} comprehensively analysed the accuracy of several available temperature and humidity sensors on the market. The findings revealed that BME280 exhibited the highest precision, even under extreme weather conditions. AM2302E showed higher errors at low temperatures, while AM2302F and SHT71 had a higher error rate within the 30-60$\%$ humidity range. Although its successor, AM2321, demonstrated a slight improvement, it remains sensitive to temperature changes. BME280 was considered the most accurate. The BME680 is the successor to the BME280, distinguished by its ability to detect TVOCs. Both sensors demonstrate similar accuracy in their measurements, with consistency in humidity, barometric pressure, and temperature. However, the BME680 exhibits a marginally higher humidity hysteresis of 1.5$\%$, in contrast to the 1.2$\%$ observed in the BME280. Furthermore, the BME680 has a reduced root mean square noise in pressure measurements, at 0.12 hPa, compared to the 0.2 hPa of the BME280~\cite{BME280,BME688}. Nevertheless, relying solely on weather sensors may prove insufficient for timely wildfire detection due to their slow rate of change.

\begin{figure}
    \centering
    \includegraphics[width=\linewidth]{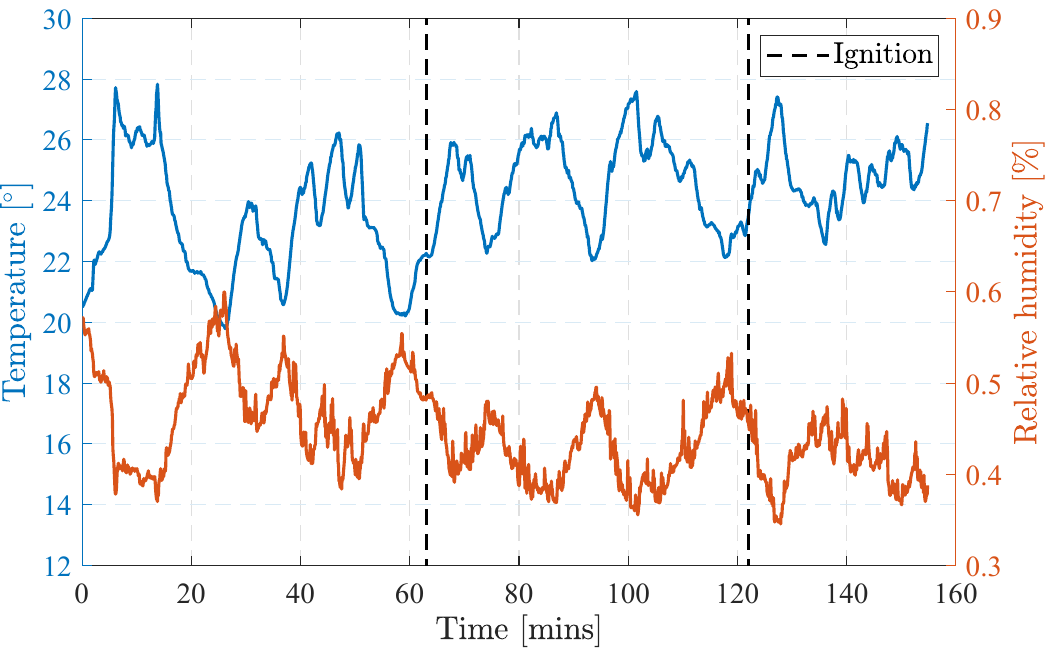}
    \caption{Average temperature and relative humidity obtained from eight BME280 sensors in an experimental outdoor burn. The results show that environmental factors have a substantial influence on temperature and weather data recorded by the sensors, overshadowing the impact of proximity to the fire.}
    \label{Fig:tempandhumid}\vspace{-1em}
\end{figure}

\begin{figure}
    \centering
    \includegraphics[width=\linewidth]{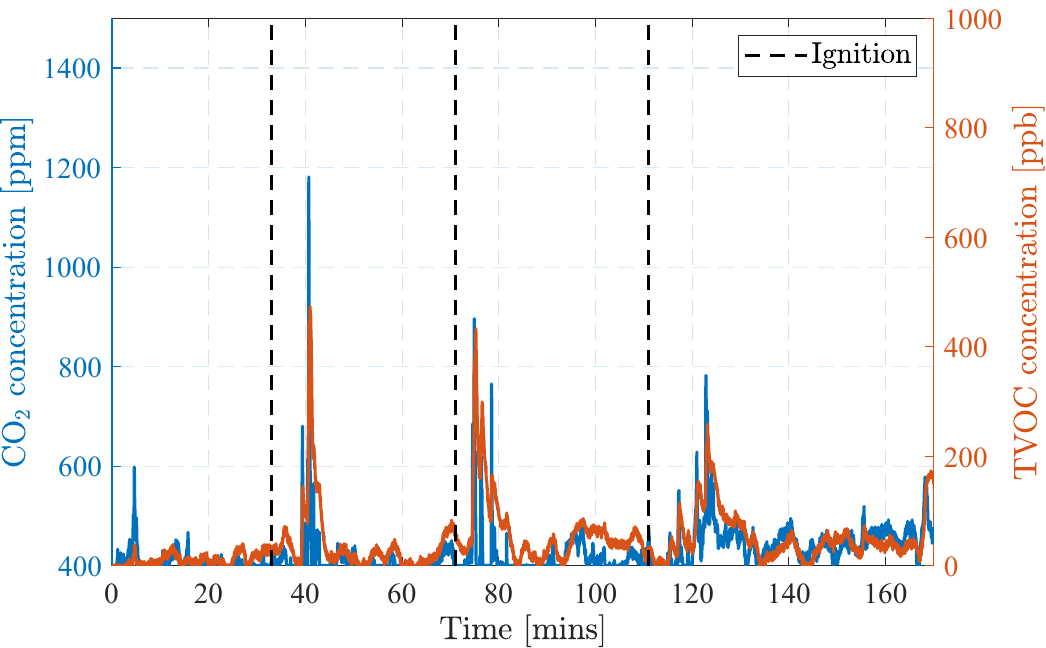}
    \caption{Measurement of the concentration of CO$_2$ and TVOC over a period of two hours, the dashed lines indicate the ignition time, distinct peaks in CO$_2$ and TVOC are observed shortly after each ignition, and these spikes occur within a period of 10 minutes.}\vspace{-1em}
    \label{Fig:CO2example}
\end{figure}

In order to evaluate the ability of weather sensors in detecting fire, our team conducted several experimental burns in a national park on various days during the low-risk wildfire season. In this set of experiments, we placed several BME280 sensors evenly spaced in a circle at varying distances, ranging from 10 to 50 meters, from the fire source. It should be noted that the BME280 sensors are highly accurate micro-sensors that can be used in IoT devices, as previously mentioned. These sensors are also used in various electronic devices, such as smart watches, smart home monitors, and other wildfire detection systems.
The temperature and relative humidity changes obtained from the sensors during a typical experimental burn are shown in Fig.~\ref{Fig:tempandhumid}. The sensors were positioned 10 metres away from the fire source. In this experimental burn, we ignited two fires using approximately 1.6 kg of \textit{Eucalyptus} leaves (equivalent to 22.535 tonnes per hectare) in a 0.7 square metered fire pit. The fires were lit at around 60 and 120 minutes after activating the sensors, each lasting ten minutes with an additional five minutes of smouldering, but we can see from Fig.~\ref{Fig:tempandhumid} that the weather sensors did not identify any notable irregularities. It suggests that depending only on weather sensors may not be sufficient to identify fires in their initial stages, particularly when the same experiment configuration is tested for gas sensors at a distance of 30~metres from the fire sources, as demonstrated in Fig.~\ref{Fig:CO2example}. However, the fires used in the example represent the lowest limit of detectable fire size. Fires up to 5 hectares can still be suppressed under mild weather conditions~\cite{10.5849/forsci.10-096}, and may alter the local temperature and humidity more than the fire in this example. Besides, as mentioned above in this subsection, fuel moisture, which is important for ignition success, can still be calculated from temperature, relative humidity and air pressure~\cite{RESCODEDIOS201564,CAWSON2022120315}. Therefore, the moisture content of the vegetation can serve as the wildfire risk indicator and provide valuable guidance on the frequency of sensor readings required for early wildfire detection.
\subsubsection{\textbf{Gas / Smoke Sensors}}
During the early stages of a wildfire, the smouldering process can result in the release of a significant amount of greenhouse gases and heat. The gases emitted by a small fire may disperse before forming a visible plume~\cite{FATEEV2017791}. Although heat can be detected by humidity, thermal and pressure sensors, a large amount of gases, such as carbon dioxide ($\mathrm{CO}_2$), methane ($\mathrm{CH}_4$) and nitrous oxide ($\mathrm{N}_2\mathrm{O}$)~\cite{GHGasEstimate}, together with particulate matter (PM), are also released from wildfires and can be measured using gas/smoke sensors. Previous research has shown that most of the carbon emitted during wildfires is in the form of $\mathrm{CO}_2$ (88$\%$), with smaller amounts of carbon monoxide (6$\%$) and organic compounds without $\mathrm{CH}_4$ (3.8$\%$)~\cite{Urbanski_2014}. As a result, gas/smoke sensors capable of detecting concentrations of these gases or smoke particles are often used in wildfire detection systems.

Various types of sensors, including metal oxide semiconductors, chemical field-effect transistors and electrochemical sensors, are currently used to measure the concentrations of total volatile organic compounds (TVOC) and CO$_2$ in the atmosphere, while optical or laser scattering sensors are commonly used to detect PM particles. These sensors provide gas concentration readings at a high sampling rate to MCUs or to the cloud server, which can facilitate early fire detection. In particular, sudden changes in gas concentrations are often more detectable than gradual changes in environmental factors such as temperature and humidity, making wildfire detection easier using such gas/smoke concentration data. We conducted another experiment with the same parameters as the one shown in Fig.~\ref{Fig:tempandhumid}, with sensors placed 30 metres from the fire and with multiple ignitions. Figure~\ref{Fig:CO2example} shows a clear correlation between fires, levels of CO$_2$, and TVOCs. The graph shows that there are multiple peaks in TVOC and CO$_2$ concentrations during the burning periods, indicating that the change in these gas concentrations due to wildfires is much more substantial than for wildfires compared to other meteorological factors such as temperature and humidity.

Despite the availability of these sensors on the market, it is still necessary to compare their characteristics and performance. Therefore, further research efforts are necessary to determine the most suitable sensor type for an IoT-based wildfire detection project. In the following sections, we present a survey of various sensor types that exhibit the potential to detect smoke PM, CO$_2$ and TVOC gases.

\begin{enumerate}[(i)]
    \item \underline{Resistive Metal Oxide Semiconductor Sensor (MoS)}: MoS sensors have gained widespread usage in a variety of IoT projects because of their cost-effectiveness and popularity in IoT development. The operational principle of MoS sensors involves the controlled heating of a metal oxide plate to a specific temperature, typically within the range of 300-400$^\circ$C. This temperature enables the metal oxide to react with the target gas, leading to a chemical reaction that reduces the electrical resistance of the semiconductor. Therefore, changes in the resistance of the MoS sensor can be used to determine the concentration of the target gas. These sensors are particularly well suited for early stage wildfire detection systems in outdoor environments, because of their affordability and ability to be deployed on a large scale. Metal oxide gas sensors exhibit robustness and stability in outdoor settings, allowing them to withstand various weather conditions, temperature fluctuations and humidity levels. As a result, they offer reliable operation for extended periods in forested areas~\cite{yu2022recent}. 
    
   Despite the advantages of MoS sensors, they are known for their relatively high energy consumption, as they require energy to heat the metal oxide plates. Additionally, extreme weather conditions, such as high humidity, can adversely affect their performance. Therefore, protective measures, such as placing sensors in weatherproof enclosures, are necessary in such conditions. MoS sensors have a high sensitivity to a wide range of gases, allowing for a quick response to the target gas, but this heightened sensitivity also introduces a high cross-sensitivity to other gases, leading to inaccurate readings, particularly in the presence of smoke. Furthermore, MoS sensors need a period of time to reach the temperature needed for precise measurements. As a result, the baseline reading of the MoS sensors can change over time due to variations in temperature, changes in humidity, and exposure to different gases, as illustrated in Fig.~\ref{Fig:idleMQ2}. Consequently, periodic calibration of the baseline is necessary when operating for extended periods. Note that other MQ series are also designed to detect various flammable gases. In this context, we have chosen the MQ2 sensor due to its wide detection range for gases emitted by wildfires, as previously mentioned in this subsection.
   
    Nowadays, advancements in integrated circuit technology have led to the integration of multiple MoS modules into a single sensor using micro-electronic mechanical systems technology. This integration enables the detection of a wider range of gases. In addition, digital control circuits have been incorporated into these sensors to facilitate baseline calibration and power management~\cite{BME688,CCS811,SGP30}. It should be noted that several studies, such as~\cite{9121967,Dryad}, have explored the application of these digital MoS sensors for fire detection. In~\cite{9121967} and \cite{jsss-2-171-2013}, it is recommended that metal oxide sensors be placed within a maximum distance of 25 metres from the fire source to ensure a timely fire detection in 15 minutes.

\begin{figure}
    \centering
    \includegraphics[width=\linewidth]{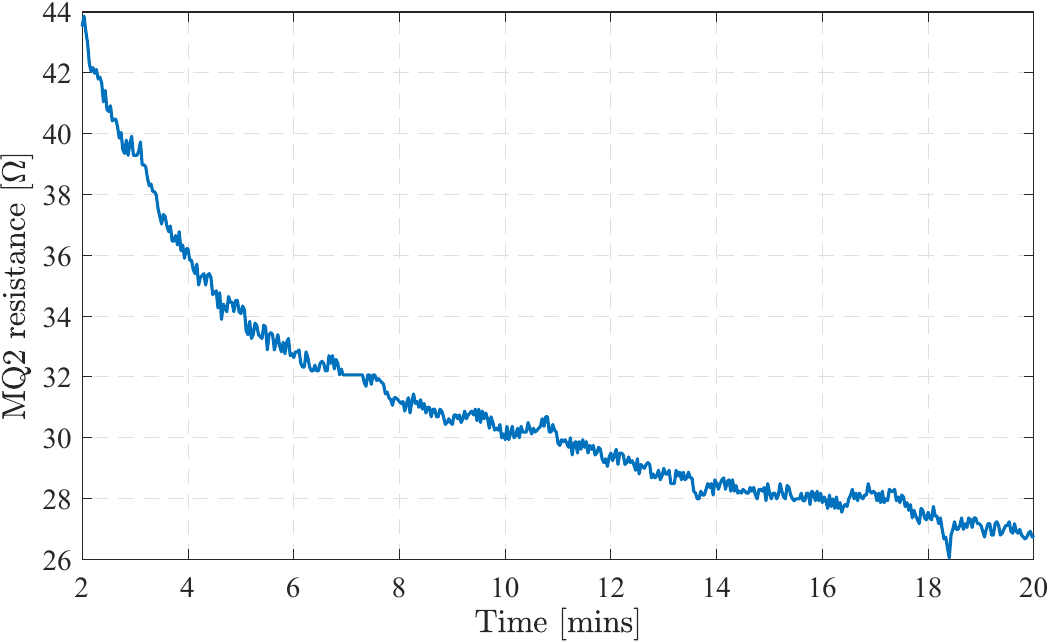}
    \caption{Measurement of a MoS sensor (MQ2) resistance in a controlled environment over 20 minutes, showing that the sensor reading baseline is drifting over time.}
    \label{Fig:idleMQ2}\vspace{-1em}
\end{figure}

    \item \underline{Catalytic Bead Sensor}: Catalytic bead sensors have been used for gas detection and monitoring for several decades as a result of their ability to respond to different sets of gases, making them highly versatile. However, with the advent of IoT projects that demand tailor-made solutions for specific objectives, the diminishing popularity of these sensors can be attributed to their cross-sensitivity issues. The catalytic bead sensor differs from the MoS sensor in its operating principle. Instead of employing a metal oxide membrane, the catalytic sensor utilises small catalyst-loaded ceramic pellets. As the target gases come into contact with the catalyst, they undergo a reaction that generates heat, which is subsequently detected by the thermistor within the sensor. The gas concentration can then be determined on the basis of the measured resistance of the thermistor.
    
    In~\cite{TYAGI20141075}, the response time and the recovery time of various metal oxide catalysts were evaluated at a distance of 6 cm from a concentration of 500 ppm of SO$_2$ and the minimum response time was found to be 80 seconds, while the minimum recovery time was 70 seconds accordingly. Catalytic sensors are more accurate than MoS sensors in detecting combustible gases such as Methane, Ethane, Butane and other VOCs. This is because catalytic sensors have low cross-sensitivity, meaning that they are less likely to respond to other gases. As a result, catalytic sensors require less calibration effort than MoS sensors.
    
    Similarly to MoS sensors, the utilisation of catalytic beads requires elevated temperatures to activate the catalytic pellets, resulting in shorter life and higher energy consumption. Furthermore, it should be noted that for IoT systems in wildfire scenarios, where most released gases are composed of CO$_2$, catalytic sensors may not be the most suitable choice for detection, as they are designed primarily for combustible gases. Furthermore, catalytic sensors are highly sensitive to temperature fluctuations and tend to be more expensive than MoS sensors. They are also prone to contamination when exposed to high concentrations of combustible gases, which can lead to degradation in sensor performance. Therefore, catalytic beads sensors are used primarily in indoor environments, such as process plants and cargo compartments of aircraft, due to their suitability for these specific applications~\cite{prashanth2020design,cleary1998evaluation}.
    
    \item \underline{Electro-Chemical Sensor}: Electrochemical sensors have been widely used in indoor fire detection and environmental monitoring projects in the past decade. However, because of its working principle, the miniaturisation of the electrochemical sensor becomes very limited and becomes less popular than the MoS sensor technology for early wildfire detection projects. Typical electrochemical sensors comprise an electrode that interfaces with an electrolyte solution, enabling the detection of target gases. When the gas is exposed, it diffuses through a porous membrane surrounding the electrode, initiating oxidisation or reduction reactions, and generating an electric current within the circuit. The gas concentration is then determined through the measured current. Compared to MoS sensors, electrochemical sensors exhibit higher selectivity toward CO$_2$ and CO due to their use of chemical reactions for gas detection, while MoS sensors are based on the physical properties of the material. Direct measurement of current induced by chemical reactions in electrochemical sensors offers selectivity higher than that of the physical properties exploited by MoS sensors, making the former more effective in detecting specific gases. Furthermore, electrochemical sensors generally have longer lifetimes and lower power consumption because their electrodes do not require heating to allow chemical reaction. However, their operating temperature range can be limited, and their sensing range may be restricted. 
    
    In \cite{doi:10.1080/15459624.2017.1388918}, a comprehensive analysis of two low-cost electrochemical sensors is presented. Although the response time of electrochemical sensors is quite promising, the accuracy of CO concentration was found to be reliable only at environmental concentrations below 12 ppm. Therefore, its application may be restricted to specific industries and air stations where low CO concentrations are expected. In general, electrochemical sensors offer improved accuracy, sensitivity and selectivity compared to MoS sensors. However, their limited operating temperature and humidity make them less suitable for low-cost IoT applications. Some examples of real-life electrochemical sensors can be found in~\cite{AlphasenseElectrochemical,4-co-2000}.
    
    \item \ul{Optical Scattering Sensor \& Non-Dispersive Infrared (NDIR) Sensor}: Recently, the use of optical or laser scattering sensors in environmental monitoring and other IoT projects has gained increased research interest because they offer higher accuracy in determining PM concentrations and faster response time when compared with MoS sensors. These laser or optical scattering sensors utilise a laser or LED beam as a light source that is directed into a detection chamber. This chamber may contain suspended particles or aerosols within the gas. As the beam traverses the chamber, it interacts with the gas/smoke particles or aerosols, causing the scattered light to disperse in various directions, including backscattering toward the detector. Positioned at an angle to the beam, a photodetector captures the scattered light, and its intensity is measured. Changes in intensity serve as an indication of the presence and concentration of gas/smoke particles or aerosols within the chamber. Through the analysis of the scattered light intensity, the concentration of PM is determined by using established calibration curves or algorithms.
    
    Analysis of the characteristics of PMs emitted during wildfire events has revealed that in the early stages of a wildfire, there is a substantial release of sawdust or dust from trees due to incomplete combustion~\cite{YIN2023117154,PREISLER2015340}. Therefore, the utilisation of optical scattering sensors also offers advantages in terms of improved accuracy and detection time, because they do not rely on chemical reactions between metal oxides or catalysts and gases. Furthermore, these sensors exhibit improved reliability compared to MoS-based sensors or other chemical reactions. However, it is important to note that optical scattering sensors can only detect the presence of obstructions in the air and cannot measure the concentration of specific gases. This limitation may impact their suitability for detecting wildfire gases, especially considering the limited travel distance of particle matter. Real-life examples supporting the efficacy of these sensors can be found in the literature, such as the work presented in~\cite{amt-11-1087-2018}, where they have been used successfully for early-stage wildfire detection. To improve the accuracy of fire detection, the authors of~\cite{benzekri2020early} implemented an IoT system that combined optical scattering sensors, NDIR sensors and electrochemical sensors.
    
    Meanwhile, there has also been growing interest among manufacturers in the NDIR sensor, leading to a wide availability of NDIR sensors for integration into IoT systems. NDIR sensors are specifically designed to detect and measure gas concentrations using the principle of infrared absorption. These sensors employ components such as an infrared light source, an optical filter and an infrared detector, similar to the photoelectric detection technique. The functionality of NDIR sensors involves transmitting infrared light through a sample chamber that contains the gas to be analysed. The target gas selectively absorbs specific wavelengths of transmitted infrared light, resulting in a reduction in the intensity of the light at a specific wavelength reaching the detector. The magnitude of light absorption is directly proportional to the concentration of the gas, enabling the NDIR sensor to accurately quantify the gas concentration. 
    
    One notable advantage of NDIR sensors is their low energy consumption, making them well suited for IoT applications. Furthermore, compared to optical scattering or other photoelectric sensors, NDIR sensors exhibit lower cross-sensitivity as a result of their reliance on infrared absorption calculations of specific gases. However, it is important to note that both NDIR and optical scattering sensors have limitations in terms of their operating temperature range, which should be taken into account. 
    
    \item \underline{Gas-Sensitive Fleid Effect Transistor (GasFET)}: GasFET sensors are relatively new to IoT projects and have not yet been widely used. However, they might become popular because of several advantages over other types of gas/smoke sensors. GasFET sensors operate on the principle of gas interactions on the sensor surface, resulting in changes in electrical conductivity. These sensors typically consist of a thin film of metal oxide deposited on a silicon substrate. The operational process of gasFET sensors involves key steps such as adsorption, conductivity modulation, gate voltage variation and signal detection~\cite{HONG2021129240,DHALL2021100116}. When the metal oxide surface is touched, gas molecules are adsorbed, potentially undergoing chemical reactions or physical interactions that modify the surface properties and affect the conductivity of the metal oxide film. This modulation in conductivity influences the electrical properties of the FET structure, which includes a gate electrode that controls the conductivity of the metal oxide channel. When a voltage is applied to the gate electrode, the conductivity of the metal oxide can be precisely controlled. The resulting changes in conductivity, caused by gas molecule adsorption, are detected as variations in the output current or voltage of the FET sensor, providing information on the concentration of the target gas in the surrounding environment. Different gases produce different responses, allowing the identification and differentiation of specific gases or volatile compounds. It is important to differentiate FET gas sensors from MoS membranes, as their underlying algorithms differ significantly.
    
    In general, gasFET sensors utilise the conductivity changes of a metal oxide thin film upon exposure to gas molecules, enabling gas detection and quantification. The design and materials of FET sensors can be optimised to provide enhanced selectivity and sensitivity to different target gases. These sensors offer advantages such as an extended operating temperature range, maintenance of performance over time, and provide consistent and accurate gas detection~\cite{DHALL2021100116}. Furthermore, FET gas sensors exhibit faster response times and greater stability compared to MoS sensors. However, they are sensitive to changes in humidity levels and can exhibit variations in baseline characteristics due to manufacturing variances between individual sensors. Furthermore, FET-type sensors are relatively newer technology compared to other sensor types. Although they offer advantages such as high sensitivity and selectivity, their development and commercialisation have been slower than those of more established sensor technologies. This slower progress may be due to complex fabrication processes and challenges in achieving consistent performance across different gas types. 
    
    Although several researchers have dedicated their efforts to advancing these sensor technologies~\cite{HONG2021129240,8614562,8956573}, it is challenging to find FET-type sensors readily available on the market for the detection of gases such as CO, CO$_2$ or TVOCs. Ongoing research and development efforts focus on the advancement of FET-type gas sensors with substantial potential for their application in various fields. The future holds promising prospects for the wide availability and utilisation of FET-type gas sensors in diverse applications.

    \item \underline{Other Gas / Smoke Sensing Techniques}: In addition to the sensing technologies mentioned above, the use of photodiode sensors has been observed in various research projects for flame detection. However, the effectiveness of fire detection by photodiode flame detectors is limited due to their maximum detection distance of only one metre. Other technologies exist that could be used for wildfire detection, such as photoionisation, chemi-luminescence, photoelectric, chromatography and resonant gas sensors. These have been evaluated to provide precise real-time gas/smoke measurements~\cite{resonantgaseg1,resonantgaseg2}, but their suitability for wildfire detection may be limited due to the substantial operational costs associated with these sensor technologies and the practical challenges involved in the deployment of these technologies in field settings.
\end{enumerate}

The performance of gas/smoke sensors can vary greatly, making it difficult for IoT designers to understand the difference in each sensor performance. To address this, many research efforts have been made to compare the accuracy, response time, selectivity and cross-sensitivity of sensors. A comparative study conducted by~\cite{GUTMACHER20111121} examined the sensitivity and accuracy of MoS, GasFET, Electro-Chemical, NDIR and optical scattering detectors. The results showed that the MoS and GasFET sensors had a sensitivity comparable to that of the other sensor types, while MoS, FET, catalytic, and electrochemical sensors display a faster response time, whereas NDIR sensors are typically slower by about 15 seconds compared to others. This observation has been further investigated by the same group in an obstructed scenario, where it was observed that the response time of the NDIR sensors could be delayed further due to blockages of the heat layers~\cite{GUTMACHER201240}. In addition, comparisons among various MoS sensors have been documented~\cite{s24041295,s21217347}. SGP30 has been observed to typically react to most VOCs faster than BME688 and ENS160, with a difference of 3--15 seconds, while ENS160 tends to produce more noisy readings. Furthermore,~\cite{s22093310} provided an extensive review of recent developments in fire detection sensors, including an explanation of their operating principles and a comparison of the performance of different types of sensors using different detection elements. 

To help IoT designers understand the features of these gas/smoke detectors, we have compiled a comparison of the performance of various types of gas/smoke sensors that are suitable for IoT projects for early wildfire detection in Table~\ref{tab:gas_sensor_comparison}. This analysis is based on commercially available sensors and takes into account factors such as cost, energy usage, reaction time, durability and precision. These aspects are beneficial to the IoT designer and can help them select the right sensor for their project. In addition, the most remarkable characteristics of the sensors are also included in the table. Furthermore, Table~\ref{tab:projectssensors} presents an overview of various IoT-based wildfire detection initiatives and the types of ground sensors used in each. It shows that many wildfire detection projects rely mainly on environmental monitoring sensors, including both weather monitoring and gas/smoke sensors, and MoS sensors are the most widely used gas/smoke sensors in these projects. In~\cite{benzekri2020early,10.1007/978-981-13-2324-9_33}, a comprehensive solution is presented, from hardware design to detection flow and DL detection, while~\cite{dampage2022forest} also offers a complete solution with implemented ML detection. \cite{ABDULKADIR20221551} provides an experimental hardware prototype, but also includes a fire-carbon spreading model which can help determine the optimal detection distance and deployment height. \cite{10.1145/1134680.1134685} presents a real project of the deployed wildfire detection system in the United States and also provides information on the sensor topology. \cite{KANAKARAJA20213907,9121967,9640900,fathima2021anintegrated,choudhary2021design,doi:10.1021/acsami.8b00245} present experimental proof-of-concepts of IoT ground sensing systems and different measurements are taken using their setup, while~\cite{fathima2021anintegrated} demonstrates the feasibility of mounting environmental monitoring sensors on UAVs to detect early wildfires. Although~\cite{forehead2020traffic} is only an example of the real deployment of the PM2.5 IoT sensing system, it also shows the ability of on board ML detection and the robustness of the use of LoRaWAN connectivity. \cite{attentis,Dryad,Waspmote} are real commercial IoT projects for early wildfire detection. Additionaly, \cite{HOEFER20121446} provides an IoT system for early fire detection featuring GaSFET sensors. In~\cite{10.1145/1134680.1134685,KANAKARAJA20213907}, both vision-based and environmental monitoring sensors were implemented on the same ED, but this approach cannot optimise the use of both technologies. To address this,~\cite{10.1145/3473037} proposed a hierarchical sensor network featuring separate camera end nodes and sensor end nodes. This network employs ground sensors to detect temperature and humidity anomalies, subsequently confirming fires using a CNN camera imaging model. Although this method may result in a relatively longer detection time compared to using vision-based or environmental sensing-based detection alone, it has demonstrated an improve accuracy to 99$\%$. Furthermore, IoT sensors assist in filtering fire alarms, allowing the camera module to remain in sleep mode most of the time, thus significantly reducing the energy consumption from camera equipped EDs.

\begin{sidewaystable}
    \caption{Performance comparison of different types of gas/smoke sensors in IoT applications}
    \label{tab:gas_sensor_comparison}
    \begin{adjustbox}{max width=\textheight}
        \begin{tabularx}{\textwidth}{l l l l l l l}
            \toprule
           \textbf{ Sensor type }& \textbf{Cost [\$]} & \textbf{Energy consumption [mW]} & \textbf{Response time [sec]} &\textbf{ Lifespan [yrs]} & \textbf{Accuracy [\%]} & \textbf{Remark} \\ 
            \midrule
            MoS (Resistive) & Low (1.5-30) & Moderate (68-90) & Moderate & 2-3 & Moderate & Cheap, high \\
            \cite{SGP30,BME688} & & & & & & cross-sensitivity\\
            Catalytic Bead & Moderate to High (70-350) & High (145-190) & Fast & 2-3 & High & Accurate,\\
            \cite{drager,Alphasense,sgxsensortech} &&&&&& volatile gases only \\
            &&&&&&\\
            Electrochemical & Low to Moderate (50+) & Low (10-60) & Fast & 2-3 & High & Fast response, \\
            \cite{AlphasenseElectrochemical,4-co-2000}&&&&&& low power cost\\
            &&&&&& limited operating\\
            &&&&&& temperatures\\
            Optical Scattering & Low to Moderate (30-100) & High (300-500) & Fast & 8-10 & High & Fast response,\\
            \cite{pms5003,sps30}&&&&&& high accuracy,\\
            &&&&&& PM particles only~\\
            &&&&&& \\
            NDIR & Low (20-50) & Low (40-300) & Moderate (60-90) & 5-10 & High & Accurate, long-life\\
            \cite{scd40,MHz14a,IRC-AT} &&&&&& slow response\\
            GasFET & N/A & Very Low ($\leq$1) & Moderate to Fast & 2-3 & High & Sensitive to humid\\
            &&&&&& environment\\
            \bottomrule
        \end{tabularx}
    \end{adjustbox}
\end{sidewaystable}

\begin{sidewaystable}
    \caption{Illustration of ground sensors employed in projects aimed at wildfire detection}
    \label{tab:projectssensors}
    \begin{adjustbox}{max width=\textheight}
        \begin{tabularx}{\textwidth}{l l l l l l l l l l l l l l l l l l }
            \toprule
            \textbf{Sensor Type} &\cite{benzekri2020early} & \cite{ABDULKADIR20221551} & \cite{10.1145/1134680.1134685} & \cite{10.1117/12.605655} & \cite{10.1007/978-981-13-2324-9_33} & \cite{9121967,9640900} & \cite{KANAKARAJA20213907} & \cite{fathima2021anintegrated} &\cite{choudhary2021design} & \cite{attentis}$^\oplus$ & \cite{Dryad}$^\oplus$ & \cite{doi:10.1021/acsami.8b00245} &\cite{dampage2022forest} &\cite{forehead2020traffic} & \cite{Waspmote}$^\oplus$ &\cite{HOEFER20121446}\\ 
            \midrule
            \textbf{Temperatures} & \checkmark & \checkmark &\checkmark & \checkmark & \checkmark & \checkmark & \checkmark &\checkmark & &\checkmark &\checkmark & & \checkmark & & \checkmark& \\
            \textbf{Humdity} &  \checkmark & \checkmark&\checkmark & &\checkmark & \checkmark & \checkmark& \checkmark &\checkmark& \checkmark & \checkmark & & \checkmark& & \checkmark&  \\
            \textbf{Pressure} &  \checkmark & & & \checkmark & & & & & &\checkmark & \checkmark & & & & \checkmark\\
            \textbf{Wind speed }& & \checkmark & & & & & & & & \checkmark & &&\\
            \textbf{Soil moisture} & & & & & & & \checkmark&\checkmark& & \checkmark & &&\\
            \textbf{MoS} & &\checkmark & & & \checkmark & \checkmark & \checkmark & &\checkmark & \checkmark & & & \checkmark & & \checkmark &\checkmark\\
            \textbf{Electro-chemcial} &  \checkmark & & & & & & & & & & & & & & \checkmark\\
            \textbf{Optical scattering} &  \checkmark & & & & & & & & & \checkmark & & & & \checkmark & \checkmark &\checkmark\\
            \textbf{NDIR} &  \checkmark & \checkmark & & & & & & & &  &\checkmark\\
            \textbf{Photodiode (flame)} & & & & & \checkmark & & \checkmark(PIR) &\checkmark & \checkmark & & & &\checkmark\\
            \textbf{FET} & & & & & & & & & & & &\checkmark &&&&\checkmark\\
            \textbf{Camera} & & &\checkmark & & & & \checkmark & & & \checkmark & & & & & &\\
            \textbf{Micro-controller} & $\dagger$ & $\ddagger$ & $\ominus$ & $\ominus$ & $\dagger$ & $\ddagger$ & ESP32 & $\ddagger$ & $\ddagger$ & N/A & N/A & $\ddagger$ & $\ddagger$ & ESP32 & $\ominus$ &  \\
            \bottomrule
        \end{tabularx}
    \end{adjustbox}
    $\dagger$ $\mathrm{Cortex-A53}$\\
    $\ddagger$ $\mathrm{ATmega328}$\\
    $\ominus$ $\mathrm{ATmega128}$\\
    $\oplus$ $\mathrm{Commerical}$
\end{sidewaystable}

\section{Data Analytic Algorithms for Wildfire Detection}\label{Sec_Algor}
Once the data collection from EDs is completed, the integration of camera images or sensor data is used to identify any fire or non-fire event. To do this, either the MCUs on the EDs or the computing unit on the central server must use data-analytic algorithms to convert the problem into a mathematical expression. The analytical process of identifying unusual events within IoT applications often aligns with the principles of anomaly detection. In the context of wildfire detection, this is specifically focused on distinguishing the rare occurrences of fires, while non-fire situations happen most of the time. Each anomaly detection technique has its own benefits and drawbacks, which make them more or less suitable for use in IoT projects.

In this section, a summary of early wildfire detection algorithms is provided. These algorithms can be classified into two main categories based on the way they obtain their detection criteria: feature detection and pattern detection in time series.  The discussion of the use of ML and DL to improve the performance of features or pattern detection for both vision-based and non-vision-based detection is also provided, as the use of ML/DL in IoT projects is gaining more research interests. Although ML/DL are heavily dependent on computers to recognise irregularities through iterative learning of features/patterns, it is essential to remember that these techniques are still based on statistical and mathematical models and can be classified as a part of the features/patterns detection. To make it easier for IoT designers in finding relevant information, we have divided this into an individual item. The following subsections discuss vision-based fire and smoke detection algorithms that use various techniques, including ML/DL and temporal/pointwise detection. Subsequently, these subsections also explore nonvision-based detection algorithms to identify anomalies either point-wise or temporally.

\subsection{Vision-Based Detection Algorithm}
Vision-based fire detection algorithms typically use image processing techniques to identify and extract relevant features from images or videos. These extracted features are then compared to the known characteristics of a fire to determine if a fire is present. Since flames and smoke are the most visible signs of fires~\cite{geetha2021machine}, most existing algorithms focus on image processing analysis to detect the features of flames and smoke. The physical properties of flames and smoke~\cite{barmpoutis2020review} can be identified using feature detection algorithms, such as colour, motion, spectral and texture. Meanwhile, pattern detection algorithms, such as spatial and temporal wavelet detection, can be used to observe the propagation of fire or smoke over a period of time. Feature detection algorithms are easier to run on IoT EDs, as they require less computing power and memory space than pattern detection algorithms. However, pattern detection algorithms used to be more accurate, as they are based on the analysis of the video stream.

\subsubsection{\textbf{Feature Detection Algorithms}}
In the following we discuss various feature detection based image processing algorithms for fire detection.

\begin{enumerate}[(i)]
    \item \underline{Colour Pixel Analysis}: The use of colour analysis for vision-based fire detection is a widely accepted approach~\cite{cappellini2005intelligent,chen2004early,dimitropoulos2012flame,celik2009fire,celik2010fast,marbach2006image,yamagishi1999fire,gati2018study,vadivu975implications}. It involves recognising the values of R, G and B of the pixels in the image to capture the colour property of flames and smoke. Different colour spaces such as YCbCr~\cite{celik2009fire}, CIELAB~\cite{celik2010fast}, YUV~\cite{marbach2006image} and HSV~\cite{yamagishi1999fire} are used to model the fire colour and also generate colour masks in the literature. RGB and HSV are the most commonly used models~\cite{barmpoutis2020review,ccetin2013video,cappellini2005intelligent,chen2004early,dimitropoulos2012flame}, with the latter providing more useful information on fire illumination. In IoT applications, the IoT camera module takes pictures of its environment and the MCU evaluates the colour values of each pixel to decide if there is a fire present. This enables the colour characteristics of fire and smoke to be recorded. Colour-pixel analysis is relatively straightforward by simply using a threshold for a variety of colour combinations in the form of adding or subtracting two colour pixel values, which is simpler to implement in IoT EDs~\cite{SHARMA2020102332}. Although colour analysis can be used to distinguish fires from images, it is not accurate enough to do so dependably, particularly when fire-coloured objects are present in the image. This results in a high rate of false alarms. To enhance precision, colour analysis is usually combined with motion analysis for fire detection. In~\cite{SHARMA2020102332}, a combination of vision-based detection and gas sensor detection is used to make a fire decision.
    \item \underline{Motion Pixel Analysis}: Motion analysis is another technique commonly used in vision-based fire detection~\cite{ccetin2013video}. This technique, known as background subtraction, captures the motion of flames and smoke by subtracting the first frame received from the subsequent frames. To achieve thresholding, in various studies, dynamic background subtraction methods based on the Gaussian mixture model (GMM) have been proposed~\cite{collins2000system,celik2007fire,xiong2007video,lee2007real,calderara2008smoke}. GMMs are used to analyse the intensity values of each pixel in the previous frames, forming a Gaussian distribution that displays the information state of each pixel~\cite{song2014background}. This image, with the background removed, can be used to identify motions, such as a fire. To detect motion pixels, a comparison between the current frame and the previous frame is necessary, which requires a bit more temporal storage for the pixels. In~\cite{celik2007fire}, it is demonstrated that fire detection using the motion pixel approach is not resistant to moving objects such as humans and animals. Therefore, relying on fire motion detection may not be reliable, as there is a chance that wildlife is moving around in the forest.
    \item \underline{Local Feature Extraction Analysis}: A range of local feature extractors can be used to improve fire detection in IoT devices, helping to recognise fire or smoke objects. The Histogram of Orientated Gradient (HOG) algorithm computes the magnitude and direction of the gradient for each pixel in a given block and then creates a histogram of the normalised values. The Scale-Invariant Feature Transform (SIFT) method identifies local features by constructing a scale-space pyramid using Gaussian masking and detecting the difference of Gaussians between adjacent pixels. Shi-Tomasi calculated the 2x2 eigenvalues of each section of the image, allowing the corners of the image to be highlighted by sorting the highest minimum of the eigenvalues. All these techniques are particularly useful for recognising the edges of an object. In~\cite{9389543}, the performance of various local feature extractors was evaluated using a Sony IMX219 camera and a Raspberry Pi 4 MCU. HOGs were found to be more successful in capturing fire and smoke characteristics. However, they were all sensitive to non-fire objects, such as humans and animals. To enhance accuracy, the MCU performed local feature extraction, and then the features were transmitted to the cloud server and used to identify fire through DL algorithms.    

    \item\underline{Other Feature Detection Techniques}: In addition to the feature detection methods discussed, there are numerous other feature extractors, such as the Harris corner detector, the Canny edge detector, binary robust invariant scalable keypoints and orientated rotated brief. These were only mentioned as they are commonly employed for fire detection on IoT devices, such as HOG and Shi-Tomasi, or can be used as a detection benchmark, such as colour and motion pixel detection and SIFT.
\end{enumerate}

\subsubsection{\textbf{Time Series Pattern Detection}}
In the following we discuss various pattern detection based image processing algorithms for fire detection.

\begin{enumerate}[(i)]
    \item \underline{Temporal Wavelet Analysis}: In addition to feature detection, more sophisticated image processing techniques can be used to detect patterns in time-series data. Temporal wavelet analysis combines colour and motion analysis to detect fires. Moving objects of fire can sometimes trigger a false alarm, but analysing the flickering property of a fire can help distinguish between real fires and these objects~\cite{ccetin2013video,hossen2018fire,marbach2006image,toreyin2006computer,chen2010multi,gunay2010fire}. The fire flickers at a frequency of approximately 10 Hz~\cite{albers1999schlieren}, and a pixel intensity oscillation frequency above 0.5 Hz suggests the presence of fire~\cite{straumann2002method}. \cite{7071763} used the Discrete Wavelet Transform (DWT) to extract information on temporal variation from a pixel by placing its intensity history in a two-stage filter bank. And by comparing the simulated results for fire pixels and non-fire pixels, fire pixels can be distinguished by counting the number of zero crossings within a certain number of frames~\cite{cetin1994signal}. Although temporal wavelet analysis combines motion and colour analysis to enhance fire detection accuracy, this comes at the cost of increased computational complexity and the need for more temporary memory capacity, leading to higher energy consumption and longer detection latency. Furthermore, smoke and flames can flicker in various ways, and to accurately capture the characteristics of early wildfires, various studies have suggested multiple approaches by integrating temporal wavelet analysis with other techniques in fire detection~\cite{7077943, ye2017effective}.

    \item \underline{Spatial-Temporal Wavelet Analysis}: In addition to temporal wavelet analysis, spatial variations can also be used to differentiate between fire and objects in motion that are coloured with fire. Fire regions in the image show a great deal of spatial variance, while objects in motion with a fire-like colour show no changes in spatial colour variations~\cite{hossen2018fire,7071763}. Spatial wavelet analysis is used to calculate the spatial variation of the image frame~\cite{ccetin2013video}. DWT is applied to decompose the original image into four subimages, which represent the horizontal, vertical and diagonal spatial energy of the image. A decision variable $\nu_4$ is calculated using these subimages, indicating the level of spatial variation of the image frame analysed. Equation $\nu_4$ can be written as~\cite{cetin1994signal},
    \begin{equation}
        \centering
        \nu_4 =\frac{1}{M N}\sum_{k,l}\left|x_{lh}\left[k,l\right]\right|^2+\left|x_{hl}\left[k,l\right]\right|^2+
    \left|x_{hh}\left[k,l\right]\right|^2
        \label{Eq1},
    \end{equation}
where the number of row and column pixels in the frame is denoted by $M$ and $N$, respectively, such that particular row and column for the calculation of the pixel value are represented by $k$ and $l$ respectively. $x_{lh}$, $x_{hl}$, $x_{hh}$ are low-high, high-low and high-high subimages, respectively. If $\nu_4$ is above a certain threshold, then the probability of the presence of a real fire is higher~\cite{cetin1994signal}. In~\cite{toreyin2006computer}, the advantage of spatial wavelet detection over colour and motion pixel detection is shown, as it significantly reduces the false alarm rate. Additionally, spatial wavelet analysis is more efficient in terms of temporal memory and detection time, since it requires only performing DWT of the image. However, the two-dimensional wavelet decomposition used in spatial wavelet analysis has a limited capacity to analyse time-series data. 
 
 To improve its ability in time-series detection, many researchers have proposed the use of spatial-temporal wavelet analysis, which combines spatial and temporal wavelet analysis and uses a weighted sum fuzzy logic for both detection methods. In~\cite{APPANA201791}, the spatial-temporal variance is used to calculate the spatial-temporal energy to detect any sudden increase in energy, which is indicative of a fire. We had also conducted a comparison of vision-based detection on a 64-bit quad-core ARM Cortex A53 with 1GB RAM, and the results demonstrate that all spatial wavelet, temporal wavelet and spatial-temporal wavelet analyses are suitable for IoT devices. The amount of time that each algorithm took to process 100 consecutive frames was 2.1, 4.1 and 6.3 minutes, respectively, while the colour and motion detection benchmark was completed in 2 minutes.
    \item \underline{Histogram of Optical Flow}: The Histogram of Optical Flow (HOF) is a popular technique for detecting spatial-temporal patterns, such as fire motion detection. To generate HOF vectors, video frames are first converted to greyscale and then optical flow is calculated using various algorithms, the most well known being the Lucas-Kanade method~\cite{10.1145/2578726.2578744}. HOG and HOF are usually used together, as they are both histogram-based descriptors. This implies that once the gradients and optical flows have been obtained, they can be quickly converted into HOG and HOF through the same process and are also computationally efficient. In addition, HOG is more likely to give details about objects, while HOF is better at recognising temporal motion. By combining HOG and HOF, many studies have shown that the accuracy of fire detection is improved compared to using either method independently~\cite{6952375,avgerinakis2012smoke}. However, all of these measurements have been performed on desktop PCs, and only one study~\cite{9389543} has demonstrated its practicality on an IoT device. The limited computing power of EDs can lead to delayed detection times when using wavelet analyses and HOF. To address this issue, researchers have proposed using location pattern detection in EDs, which transmit data for processing by ML or DL algorithms~\cite{9389543,HUANG2022104737}.
    
    \item \underline{Other Time Series Pattern Detection Algorithms}: In addition to the time-series pattern detection methods mentioned, there are other detection algorithms such as auto-regression and spectral analysis, while the most commonly used and widely adopted time-series detection model for capturing fire and smoke in IoT EDs is listed.
\end{enumerate}

\subsubsection{\textbf{Machine Learning and Deep Learning Algorithms}}
In the following we discuss
various ML and DL based detection algorithms
for fire detection.

\begin{enumerate}[(i)]
    \item ML algorithms have been increasingly utilized in fire detection tasks in recent years~\cite{app12052646,8451657,9356284,8022860,luo2018fire,10.1145/3319921.3319962,8385121,iandola2016squeezenet,jadon2019firenet,he2015deep}. Simple ML methods can be applied directly to enhance feature or texture detection, such as colour and motion pixels. In~\cite{2016.0193}, it shows that the use of a support vector machine (SVM) can improve the accuracy of forest fire detection by colour pixel analysis by 0.5\%. In~\cite{wevj13020023}, K-means algorithm was used to re-cluster the pre-labelled anchor frame parts in the dataset. Furthermore, the author suggested that using K-means to cluster the fire-like area into local binary patterns for SVM could improve the accuracy by more than 1. 75\%. In addition to feature detection, ML methods can also help to make decisions from different detection algorithms. In~\cite{9389543}, fire detection is achieved by combining classification results using a linear SVM, and the performance of this ML detection outperforms the simple temporal wavelet detection of 5\%.
\newline

    \item Another popular method is using Convolution Neural Networks~(CNN) in DL to train a classification model to automatically identify fire and non-fire images. Unlike conventional algorithms that require hand-crafted features, DL algorithms can effectively learn the features themselves, resulting in better performance~\cite{geetha2021machine}. A typical CNN consists of an input layer, an output layer, multiple common layers and hidden layers, which are usually convolutional layers. Convolution is the main process for CNNs to learn features, completed by multiplication of elements between the convolution matrix and the input image. The convolutional process generates values that represent information about the features of the input image, which are then passed through common layers such as activation, pooling, normalisation and fully connected layers~\cite{gaur2020video}. The most important features are highlighted during the pooling process. The last few layers are always fully connected layers, where all the information on different layers of CNN is combined to obtain a final result. The final result is compared with the input data benchmark to generate a numerical indication of how well the model predicts the input. This indication is passed through each CNN layer from end to end to correct for the weights associated with each layer using backpropagation~(BP). CNNs not only represent the types of feature that the model focuses on but also define the CNN itself. Different CNNs perform different tasks with different weights in the layers, even with the same layer structure. AlexNet, ResNet, GoogLeNet and VGG-Net are the most popular CNNs used to solve computer vision problems~\cite{10.1145/3065386,he2015deep,he2016identity,Szegedy_2015_CVPR,simonyan2015deep}.

    In DL, fire detection is a binary classification problem that requires a binary classifier to solve the problem. The complexity of the problem can be reduced by focusing on wildfire detection with binary classifiers. An overview of current lightweight CNNs that are built for fire detection and are possible to implement on a resource-constrained device is given below. In~\cite{app12052646},  the classifier uses a feature learning method that learns basic low-level structures of fire first and high-level semantic features about fire later. Their classifier was compared with other existing fire detection classifiers, including FireNet, InceptionNetV1-OnFire, NasNet-A-OnFire and ShuffleNetV2-OnFire. InceptionNetV1-OnFire achieves the highest accuracy of 95.6$\%$ with 12 fps, and NasNet-A-OnFire can achieve the lowest fps of 7 with 95.3$\%$ accuracy. In~\cite{8022860}, the authors proposed a 14-layer deep CNN model capable of automatic feature extraction and classification in fire smoke, achieving a detection rate of 96.37$\%$ and a false alarm rate of 0.60\%. In~\cite{luo2018fire}, a new lightweight smoke detection CNN is trained using pre-processed images and achieves a detection rate of 99.8\%, a false alarm rate of 0.31\%, and an overall accuracy of 99.7\%. In \cite{10.1145/3319921.3319962}, a fire detection approach was proposed that combines the strengths of conventional and DL algorithms and can detect both flames and smoke in the image, achieving good results.  To achieve fire image detection using DL in power-limited devices,  a lightweight CNN architecture is proposed in~\cite{8385121}. Fire detection and localisation can be achieved using the proposed CNN architecture, employing techniques such as small kernels, transfer learning, model tuning and a selection algorithm for feature maps. As a result, the size of the architecture is reduced from 238 MB to 3 MB. Additionally, valuable information is provided on the identification of burning objects and the precise location of the fire within the image frame. However, a more efficient alternative approach is demonstrated by the introduction of a scratch-based model design that uses small CNNs. Consequently, this research has led to the development of the well-known CNN model, Firenet, which can be seamlessly integrated into power-limited devices such as the Raspberry Pi~\cite{jadon2019firenet}.
    
    Dataset also plays a crucial role in achieving high performance in DLs. To train a fire classifier, a data set consisting of annotated positive and negative images is required. Videos can also be used as a valid dataset if they are converted to images. In~\cite{Zhang2016/01}, researchers achieved a test accuracy of 90\% by training their model in online fire videos, while another study achieved an accuracy of 97\% with a diverse database that included self-recorded videos and various online platforms~\cite{yar2021vision}. Public datasets available for fire detection research include the Fire-net dataset with videos and images, the Fire Flame Dataset with 3000 images classified into fire, smoke, neutral and the VisiFire dataset that offers video clips for fire and smoke detection~\cite{deepquestai,VisiFire}. Recently, there has been a notable trend among global communities toward collaborative exchange of experimental and prescribed burns footages. This initiative aims to improve fire detection algorithms around the world by providing access to extensive and varied datasets. A notable example of such an effort is documented in~\cite{dewangan2022figlib}.
\end{enumerate}
Table~\ref{tab:algorithmcamera} presents a summary of the vision-based methods utilised by various IoT wildfires detection systems, detailing the advantages and disadvantages of these methods as discussed in this section. Although the use of vision-based technologies has shown great potential in the analysis of satellite images, real-time camera systems on UAVs, and alleviating the workload previously carried out by watchtower observers responsible for detecting occurrences of wildfires~\cite{aus_satellite}, it remains susceptible to interference from various environmental factors, such as weather conditions (e.g., cloud cover, rainfall and reflection of sunlight), as well as physical obstructions such as trees and buildings. As a result, the response time to imaging detection can be significantly delayed when camera modules are used on IoT EDs. Besides, if the number of cameras mounted on an IoT ED is limited, the camera is usually required to rotate and scan every hour to cover a full 360 degrees of view, which will increase the time it takes to detect something. Furthermore, these camera systems can be expensive to deploy and consume significant amounts of energy. Therefore, early wildfire detection using environmental sensing data in IoT systems has been explored as a potential alternative.

\subsection{Environmental Sensing-Based Detection Algorithm} 
In addition to vision-based detection methods, early wildfire detection can also be achieved using data from environmental monitoring sensors. MCUs can apply various detection algorithms to detect abnormal readings, which could indicate the presence of smoke or fire. This process is commonly referred to as anomaly detection. It is essential that the data given to the detection algorithm have considerable changes when a fire occurs, so that it can differentiate between normal and abnormal data patterns. And there are multiple ways to detect abnormalities, ranging from simple statistical techniques to advanced DL methods. These techniques exhibit diverse capabilities in processing input, from singular data points to continuous data streams within specific temporal windows. Using ML/DL, anomaly detection methodologies can be further differentiated into unsupervised and supervised frameworks. To reflect the significance of anomaly detection being used in early wildfire detection, we divide these techniques into three categories: threshold-based anomaly detection (point-wise), which focuses on individual data points, anomaly detection in time series (pattern-wise), which examines observations within a specified short-term period, and ML/DL strategies for early wildfire detection due to their increasing use in anomaly detection.

\begin{table}
    \centering
        \caption{Illustration of some well-known vision-based techniques employed in wildfire detection}
    \scalebox{.725}[.725]{\begin{tabular}{l l l l}
    \toprule
    \textbf{Algorithm}& \textbf{Reference(s)}	 & \textbf{Pros}& \textbf{Cons}\\
    \midrule
\multirow{2}{*}{Feature Detection} &	~\cite{cappellini2005intelligent,chen2004early,dimitropoulos2012flame,celik2009fire,celik2010fast,marbach2006image,yamagishi1999fire,gati2018study,vadivu975implications} &	\multirow{2}{*}{Easy to deploy} & Cannot detect temporal  \\
&~\cite{collins2000system,celik2007fire,xiong2007video,lee2007real,calderara2008smoke} & & smoke pattern \\

Time Series Detection & \multirow{2}{*}{\cite{hossen2018fire,toreyin2006computer,chen2010multi,gunay2010fire, albers1999schlieren, straumann2002method, 7071763,cetin1994signal, 7077943, ye2017effective}} & Temporal pattern & Longer\\ 
(Temporal Detection) & & can be extracted &processing time \\

\multirow{2}{*}{DL Neural Networks} & \multirow{2}{*}{\cite{10.1145/3065386,he2015deep,he2016identity,Szegedy_2015_CVPR,simonyan2015deep}}&  Highly accurate  & Resource requirement \\
& & and well-established & and delay response \\
         \bottomrule
    \end{tabular}}
    \label{tab:algorithmcamera}\vspace{-2em}
\end{table}

\subsubsection{\textbf{Threshold-Based Anomaly Detection}}
In order to detect and respond to abnormal readings, a commonly used approach is the implementation of a threshold limit. This method allows users to define a specific threshold value for recorded readings. If the readings exceed this predetermined threshold, it indicates that the readings have exceeded the normal range and, subsequently, trigger the activation of the fire alarm ($\lambda$). The corresponding equation that represents this relationship is as follows,
\begin{equation}
    \centering
    \lambda(t) = \left(\land_{i=1}^{n}\left(x_i(t)>X_{i\mathrm{th}}\right)\right) \land \left(\lor_{j=1}^{m}\left(x_j(t)>X_{j\mathrm{th}}\right)\right),
    \label{thresholdequation}
\end{equation}
where $\lambda(t)$ is true if all sensors in a certain subset have readings $x_i(t)$  at time $t$ exceeding their respective thresholds $X_{i\mathrm{th}}$ $\forall~i\in n$ and if any sensor from a different subset has a reading $x_j(t)$ exceeding its threshold $X_{j\mathrm{th}}$ $\forall~j\in m$. This setup allows for a general expression that accommodates various sensor data configurations, incorporating conditions that must all be met simultaneously and those where meeting any one condition is sufficient. The most common way to determine $\mathrm{X_{th}}$ is by using the z-score. This involves finding an appropriate threshold value, which is expressed as $\mathrm{X_{th}}=\mu\pm z\frac{\sigma}{\sqrt{n}}$. The sample mean, $\mu$, is calculated from either non-fire state data or fire state data. The confidence level is indicated by the parameter $z$. The standard deviation of the sample is represented by $\sigma$, while $n$ is the sample size. This approach assumes that the distribution of sensor data follows a Gaussian distribution. The z-score can then be applied, which states that approximately 99.73\% of all samples will be within 3$\frac{\sigma}{\sqrt{n}}$ of the mean. The desired level of sensitivity can be adjusted considering whether a particular sample falls within the range 95. 45\% (corresponding to 2$\frac{\sigma}{\sqrt{n}}$) or 68.27\% (corresponding to 1$\frac{\sigma}{\sqrt{n}}$) range of all samples~\cite{10.1145/2783258.2788611}.

Additionally, the Local Outlier Factor (LOF) is also a prominent algorithm used for anomaly detection, where it assesses the outlying level for each data point within a dataset by comparing its local density with that of its neighbouring data points. In some of the literature, the LOF is classified as an ML algorithm. However, due to its simplicity, we classified it as a threshold-based detection method here. The LOF equation, which is well-established, is expressed as follows,
\begin{equation}
    LOF(x) = \frac{\sum_{o\in \mathrm{N_k}(x)}\frac{\mathrm{LRD_k}(o)}{\mathrm{LRD_k}(x)}}{\left|\mathrm{N_k}(x)\right|}
\end{equation}
\begin{equation}
    \mathrm{LRD_k}(x) =\frac{\left|\mathrm{N_k}(x)\right|}{\sum_{o\in \mathrm{N_k}(x)}||x,o||},
\end{equation}
where $||x,o||$ is the reachable distance between point $x$ and its neighbour at point $o$, $\left|\mathrm{N_k}(x)\right|$ represents the count of nearest neighbours and LRD denotes the local reachable density, which can be interpreted as the inverse of the average reachable distance to its closest k neighbours. If the LOF is approximately 1 or less, it can be inferred that the data point is within the normal range. In contrast, if the LOF exceeds 1, it indicates the presence of an outlier, thus identifying an anomaly. Furthermore, additional statistical techniques, including moving averages and exponential smoothing, can be integrated into z-score calculation and LOF analysis. This involves comparing the observed value with the moving or exponential average to identify notable deviations and signify potential anomalies~\cite{10.1145/3338840.3355641}. Generally, statistical techniques are the most straightforward methods for early fire detection. By using pre-recorded background readings, it is possible to identify a threshold value without the need for on-board processing. This makes it the most energy-efficient option and commonly used detection model for wildfire detection. However, its accuracy is limited by threshold values, which makes it less precise than other detection methods. Therefore, many research projects employ statistical thresholds to perform initial filtering before performing ML/DL algorithms that require high computing power~\cite{5690033,9640900,ghuge2021forest,dampage2022forest,forehead2020traffic,9530090,9332990}.

\subsubsection{\textbf{Time Series Detection}}
In additional to the threshold-based point-wise detection, the use of detecting changes in power density over a short period of time and seasonality detection has been extensively studied as an alternative, particularly for anomaly detection in IoT systems.

Various techniques have been explored for the detection of abnormal changes over a period of time, including Bayesian online changepoint detection~\cite{adams2007bayesian}, cumulative sum control chart (CUSUM)~\cite{xie2023window} and bottom-up segmentation~\cite{w13121633}. In general, the primary objective of change-point anomaly detection is to identify significant changes or intervals in a time series that deviate from the underlying behaviour or pattern. This involves segmenting the time series, fitting statistical models to each segment and detecting changes between adjacent segments. The presence of an anomaly is inferred if a significant change is detected. The effectiveness of change point detection depends on segmentation, model selection and assumptions made about the data. These approaches capture various types of anomalies, including sudden changes, gradual changes and recurring patterns. In~\cite{10006406}, CUSUM was used to detect unusual body movements on a low-power ARM-Cortex R4F. In addition, the implementation of CUSUM for anomaly detection on fog platforms was demonstrated in multiple projects, as summarised in~\cite{ALENCAR202077}. A sequential approach for anomaly detection using Bayesian changepoint detection in the IoT is also proposed in~\cite{ELKHATIB2020222}. Numerous projects have demonstrated the effectiveness of employing changepoint detection models in anomaly detection, particularly within on-board IoT systems. Although changepoint detection exhibits a higher level of complexity compared to point-wise detection techniques, it remains less complex than advanced seasonality or ML/DL detection methods. Moreover, changepoint detection has been observed to offer superior accuracy in anomaly detection in contrast to point-wise detection approaches.

An alternative approach to analysing time-series data is the seasonality detection method, such as seasonal decomposition methods and seasonal forecasting methods. In seasonality detection, the trend or seasonal component captures the overall direction of the data over a long-term period, providing information on whether the time series is increasing, decreasing, or relatively stable. Seasonal decomposition techniques are used to identify patterns in the data that repeat at fixed intervals by isolating the seasonal component so that any recurring patterns in the data can be retrieved. These fluctuations can be considered anomalies or unexpected behaviour in the data~\cite{hochenbaum2017automatic}. In contrast, seasonality forecasting techniques, such as Holt-Winters and Fourier analysis, also consider seasonality when predicting. However, to enable anomaly detection, users still need to design their own anomaly-checking mechanism. A straightforward detection algorithm is to calculate the mean square error and feed it into a given threshold and confidence intervals to determine if it is an error. In~\cite{ansari2023intelligent}, the author evaluated the effectiveness of various IoT air quality prediction models using seasonal forecasting methods and seasonality ML/DL prediction models. The results of the study showed that it is possible to predict seasonality in EDs using low computing power Cortex-M0 or high-performance Cortex A53. However, seasonal anomaly detection models are not widely used in anomaly detection due to the fact that the full data set may not be available if the seasonal patterns last longer than one year. For example, in~\cite{10225693}, researchers have shown that Holt-Winters requires an additional 40KB of data to correctly predict CO$_2$ and the temperature level on an ESP32. Besides, these methods assume that the time series has repeated seasonal and stationary patterns. If the data includes irregular or non-recurring seasonal patterns or if the seasonal patterns change over time, these methods may not be able to accurately capture the underlying structure. 

\subsubsection{\textbf{Anomaly Detection Using ML and DL}}
In addition to the detection methods mentioned, the use of ML or DL for anomaly detection has been increasing due to its ability to capture complex relationships and non-linear patterns in data, as well as its adaptability and scalability in dealing with different input features. These ML/DL approaches make it possible to process large or multi-dimensional datasets quickly and easily for point-wise or time-series detection. Moreover, ML techniques can be applied to various components of a network, from IoT devices to network servers. These advantageous characteristics have led to the growing adoption of ML for anomaly detection. Several basic supervised ML algorithms have been used for anomaly detection, such as SVM, K-nearest neighbours (KNN) and decision tree (DT)~\cite{CHEN2022108046}. SVM uses hyperplanes to classify data into normal or anomaly groups, while KNN determines anomalies based on the Euclidean distance to their nearest-neighbour group. DT uses reduction of entropy values to identify relevant information for anomaly detection. However, these ML techniques require user-labelled data points, which are usually obtained by manual detection. For instance, when it comes to fire detection, users must label the time stream of sensor readings as either fire or non-fire states. The accuracy of these methods is highly dependent on the accuracy of the labelling process. Therefore, some anomaly detection methods are developed using ML clustering techniques, such as K-means and GMMs, and many IoT projects have implemented these approaches, as shown in~\cite{dampage2022forest,7794093}. Furthermore, some ML forecasting models, such as Autoregressive Integrated Moving Average (ARIMA) and Seasonal Autoregressive Integrated Moving Average (SARIMA), can also be combined with statistical outlier methods to detect anomalies without supervised learning~\cite{panuju2010historical}. If the input dimensions of ML are small, the energy consumption of ML can be comparable to that of statistical methods. However, the greatest advantages of ML are its ability to recognise interconnections between different inputs and to detect anomalies. Furthermore, its complexity and computing power requirements are lower than those of the DL methods, making it more suitable for deployment on resource-constrained EDs. Recent IoT ground sensing system projects have widely used ML anomaly detection algorithms, as summarised in Table~\ref{tab:algorithmIoTprj}.

To improve the precision of anomaly detection and reduce the likelihood of errors caused by human labelling in supervised ML methods, DL approaches have gained popularity. DL algorithms are capable of modelling complex nonlinear relationships between input features and output labels, allowing for the detection of anomalies that may be missed by simpler linear models. DL models such as Artificial Neural Networks (ANNs) and CNNs can leverage networks that have been pre-trained on large and diverse datasets from similar domains. This can reduce the need for a large number of labelled data during the initial training phase and can also improve the recognition of complex patterns and features in the data. When using ANNs for anomaly detection, the raw input data are processed through multiple hidden layers, which work to identify and capture representations that are indicative of anomalous behaviour. Subsequently, the output layer produces a prediction or score that indicates whether the data are anomalous or normal. In~\cite{WANG2013413}, three ANN models were used, namely BP, radial basis function (RBF) and probabilistic neural network, for indoor fire detection using weather and gas sensors. The results demonstrated the exceptional performance of all three models in the detection of fire incidents. In addition to ANN, recurrent neural networks (RNNs) were developed to process sequential data. These networks have recurrent connections that allow them to store information from previous steps or moments. This allows RNNs to recognise and learn patterns in sequences, making them more effective than other DL approaches, since the input data are likely to be a time series of environmental readings from sensors. In particular, long-short-term memory networks (LSTMs), a specific type of RNN, were developed to address the problem of vanishing gradients faced by traditional RNNs. LSTM networks introduce a specialised memory cell that can retain information over long periods of time. This makes it more effective in capturing long-term dependencies in sequential data~\cite{benzekri2020early}. Furthermore, LSTM networks and other DL models, such as autoencoders and multilayer perceptrons, can be used for unsupervised or self-supervised learning to reduce errors. These methods are simpler than other self-supervised DL clustering models, making them easier to implement on the edge with relatively low energy consumption. Additionally, these DL models can be trained online or in real time, making them more adaptive to changing environmental conditions, especially useful for early wildfire detection. Some newly developed DL models for unsupervised time series anomaly detection are also reported in~\cite{xu2022calibrated,wang2022flowadgan}. However, their compatibility with performance-constrained EDs has not been investigated yet. The study conducted by~\cite{benzekri2020early} investigates the performance of three RNNs, namely simple RNN, LSTM and GRU-RNN, for early forest fire detection. The results demonstrate that all models achieve an accuracy that exceeds 99.7\%. Specifically, the GRU-RNN model exhibits the highest accuracy and demonstrates reduced complexity compared to LSTM. These findings highlight the effectiveness of RNN models in accurately detecting forest fires at an early stage, and the GRU-RNN model shows potential advantages in terms of accuracy and complexity. 

Table.~\ref{tab:algorithmIoTprj} presents an summary of the anomaly detection algorithms used in recent forest fire detection projects. In general, DL and ML approaches are widely used in these projects. Most of these initiatives employ a threshold limit, based on the z-score, regression, or LOF, for initial filtering. Upon triggering consecutive alerts, the ED transmits data to the server to perform fire detection or performs specific fire detection algorithms on the ED itself. Note that in this section, we have focused on a few anomaly detection techniques that are suitable for implementation on EDs or the gateway for wildfire detection. The selection of these methods was based on their computing power requirements, complexity, and whether they have been previously implemented.

\begin{table}
    \centering
        \caption{Illustration of well-known anomaly detection algorithms employed in wildfire detection}
    \scalebox{.7125}[.7125]{\begin{tabular}{l l l l}
    \toprule
    \textbf{Algorithm}& \textbf{Reference(s)}	 & \textbf{Pros}& \textbf{Cons}\\
    \midrule
\multirow{2}{*}{Statistical Threshold} &	\cite{9640900,dampage2022forest,forehead2020traffic} &	\multirow{2}{*}{Easy to deploy} & \multirow{2}{*}{Inaccurate}  \\
&\cite{7794093,9530090,5690033} & & \\
\multirow{2}{*}{Time Series Detection} &\multirow{2}{*}{\cite{panuju2010historical}} & \multirow{2}{*}{Pattern-wise detection} & Complexity in\\ &&&multivariate data \\
\multirow{2}{*}{ML Classfication} &	\cite{fathima2021anintegrated,dampage2022forest,forehead2020traffic,CHEN2022108046} & Efficient multivariate &	Require periodic \\
&\cite{7794093,4640996,9530253}& data integration & re-training of models\\
\multirow{2}{*}{DL Neural Networks} & \cite{benzekri2020early,5690033,9530090,9332990} &\multirow{2}{*}{Accurate} & Resource requirement \\
&\cite{s19143150,1244476,panuju2010historical,9998196,4640996}&& delay response \\
         \bottomrule
    \end{tabular}}
    \label{tab:algorithmIoTprj}\vspace{-2em}
\end{table}

\section{Operational Challenges}\label{Sec_Challenges}
In order to establish an IoT ground sensing system aimed at early fire detection,  a holistic consideration of various factors is essential. This extends beyond the implementation of sensors, gateways and servers, encompassing aspects such as communication technologies, medium access control (MAC) layer control, energy consumption, detection latency and device sustainability. The purpose of this section is to discuss the challenges and constraints related to the implementation of IoT sensing systems to offer valuable perspectives to individuals considering designing their own early wildfire detection system.

\subsection{Sensors}
As mentioned in Sec.~\ref{Sec_lit_review_new}, gas/smoke sensors have different characteristics when it comes to wildfire detection. Some sensors have minimal cross-sensitivity, but their operating temperature and humidity ranges are limited. On tperformr hand, some sensors are highly accurate, but they take time to precisely measure the concentration of the target gases. Additionally, the cost and availability of these sensors, such as energy consumption, maintenance cost and size, are crucial considerations. In fact, numerous MoS sensors require different voltage levels to operate, such as the SGP30 and CCS811 which run at 1.8 volts. Using a linear voltage regulator on a 3.3 volt battery can lead to significant energy loss, and therefore a buck converter might be required to efficiently convert DC-to-DC voltage for these sensors. Although~\cite{li2004motion} indicates that the detection of fire flicker characteristics is not influenced by the types of combustible materials or the distance from the source, employing a typical IoT camera with a 62.2$^\circ$ horizontal and 48.8$^\circ$ vertical field of view to identify fires at operational distances up to 500 meters within a forest presents significant challenges. For example, to detect a fire from 500 meters away with a minimum required resolution of 16x16 pixels for reliable detection, the fire must be at least 5.03 meters wide and 6.72 meters tall, as per field of view calculations. In addition, it is essential to take into account not only the cost of the sensors, but also the operational costs associated with the equipment and needed for their operation and maintenance. A thorough assessment of these elements is necessary when selecting the right sensors for a particular purpose. Sensors can be delicate and need to be calibrated regularly, which can make them costly to use in a wildfire detection system. They may be exposed to extreme environmental conditions and may need to be replaced often. Besides, the pthat use edge computing.n of these sensors in the field also presents several challenges, including the establishment of a secure mounting site and the protection of the sensors against environmental hazards. These challenges pose obstacles in deploying sensors in remote regions, which are often the areas most vulnerable to wildfires.

\subsection{Communication link and Access Layer Control}\label{Sec_MAC}
In addition, the reliability and robustness of communication within sensor systems pose significant challenges. One specific challenge arises from the dependence on a clear line-of-sight between transmitters and receivers to establish optimal communication links. However, in IoT ground sensing systems, especially those used for wildfire detection, non-line-of-sight conditions are common due to obstructions such as forests and smokey environments containing high levels of PM2.5 and PM10 particles. To reduce the adverse impacts of short-wave radio technologies and address the issues caused by blocked communication, it is essential to perform a comprehensive analysis of this phenomenon to understand amplified channel pathloss and slow-fading attenuation so that transmission power can be adjusted to reduce these effects. Pathloss measurements reported in \cite{7247253} reveal that a 922 MHz signal can experience pathloss of up to 100 dB at a distance of 200 metres, projected to rise to 120 dB at 400 metres and even higher at 2.4 GHz frequencies. Furthermore, in the IoT system for early wildfire detection, which is often deployed in remote areas with limited coverage of the cellular network, it becomes imperative to establish extended-range communication between end devices and gateways to facilitate data transmission to the server. Consequently, the transmit power of the end devices needs to be enhanced, resulting in increased energy consumption. To address these challenges, the adoption of an LPWAN protocol suitable for extended-range communication becomes indispensable.

\begin{figure}
    \centering
    \includegraphics[width=\linewidth]{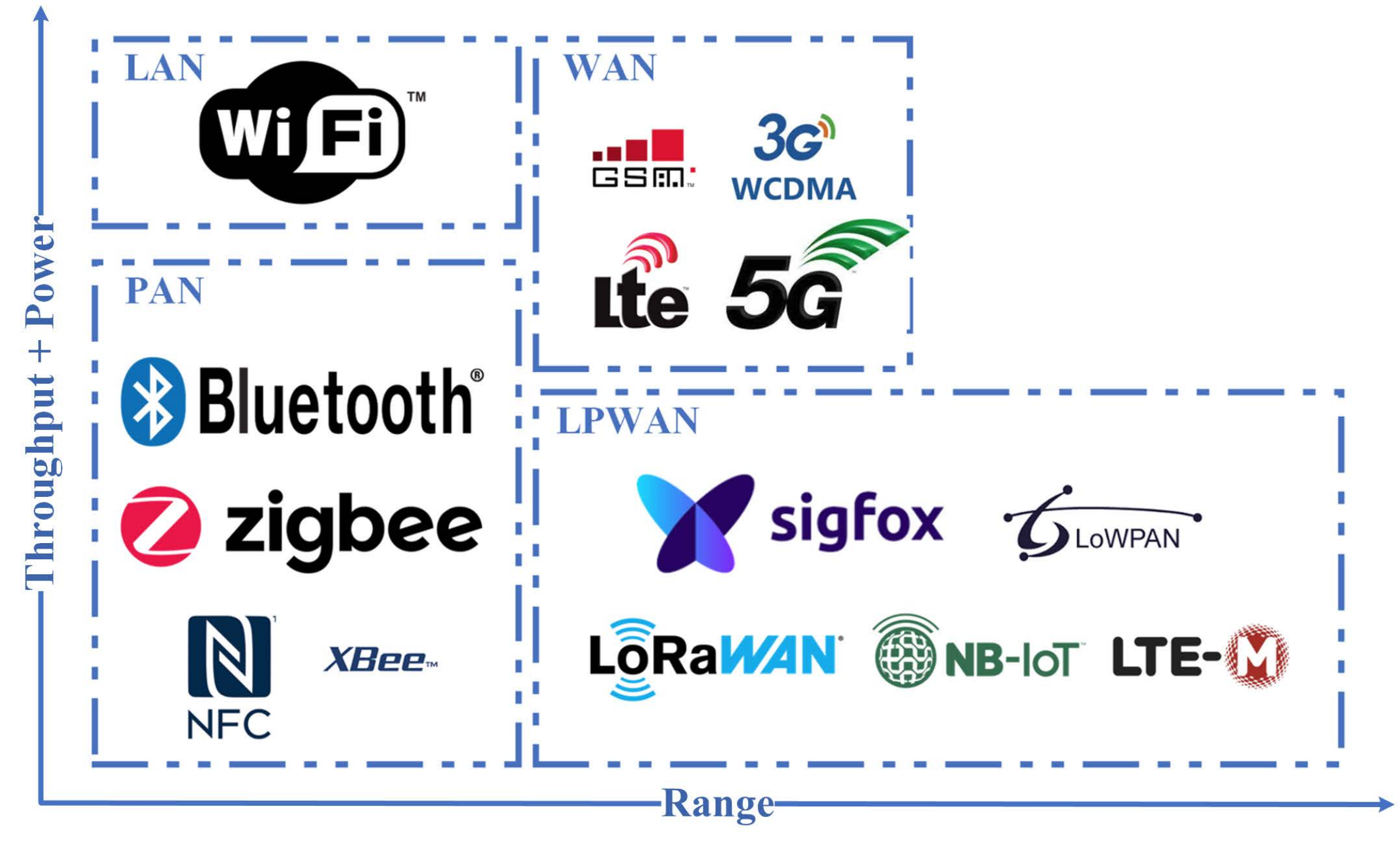}
    \caption{Illustration of LPWAN performance compared with other wireless technologies.}
    \label{fig:LPWANtech}\vspace{-1em}
\end{figure}

Typically, IoT systems employ LPWAN protocols for communication between the end nodes and the gateway. These LPWAN communication protocols are characterised by ultra-narrowband or simple modulation/demodulation techniques, enabling low energy consumption during communication. The use of ultra-narrowband signals reduces the chance of packet collision, while wideband signals with distinctive patterns allow for easy differentiation from other signals, facilitating extended-range communication. Key players in the LPWAN domain include LoRaWAN, Sigfox, 6LoWPAN, NBIoT and LTE-M~\cite{mekki2018overview}. Although NBIoT and LTE-M operate within licensed spectrum bands, LoRaWAN together with Sigfox and 6LoWPAN utilise ISM-shared bands. A performance analysis comparing various wireless communication technologies, as shown in Fig.~\ref{fig:LPWANtech}, demonstrates that LPWAN offers the longest range and the lowest energy consumption. However, it should be noted that LPWAN technologies exhibit the lowest data throughput. As a result, IoT system designers must prioritise the significance of data and determine which data to transmit and discard. Although most LPWAN technologies support packet fragmentation, this feature can exaggerate the likelihood of packet collisions and the overall packet loss rate within the system.

In wireless communication, packet collision occurs when multiple packets are simultaneously transmitted to a receiver, making it impossible to differentiate between individual packets. In the absence of proper MAC layer control, IoT packets follow a random access communication model, with the packet successful rate denoted by $e^{-2G}$, where $G$ is the traffic load measured by the total number of transmission attempts per frame time. If we assumed that all EDs operate at identical frame times within the system and transmit at uniform intervals, the limiting factor to achieve optimal packet success is defined by the expression $N\times A_t\times T_p = G = 0.5$, in accordance with the peak throughput characteristic of the ALOHA protocol, where $N$ represents the quantity of EDs, $A_t$ denotes the attempt rate per device, and $T_p$ signifies the frame time. To mitigate this issue, an MAC layer is responsible for regulating channel traffic, thereby ensuring successful packet reception. The MAC layer in most LPWAN technologies is typically predefined, and some employ random access or Listen Before Transmit~(LBT) techniques~\cite{sigfox,LoRa}, while NB-IoT and LTE-M use the same MAC control as cellular networks due to their integration in the cellular communication system. \cite{8766700} explores a modified LBT-type MAC control within LoRaWAN, demonstrating its ability to significantly reduce packet collisions and thus enhance support for numerous IoT EDs. Similarly,~\cite{8693695} presents simulation findings indicating that the average delay for LBT-LoRa is approximately 7 seconds compared to random access methods, further affirming the suitability of LBT for IoT communications. Another type of MAC layer control can be found in~\cite{electronics8121435}, where a custom Carrier-sense multiple access with collision avoidance (CSMA/CA) MAC layer was implemented in the LoRa communication system to improve the packet success rate. Nevertheless, the adoption of an in-house MAC layer control can lead to increased energy consumption for end devices due to requirements such as channel listening and synchronisation. An alternative approach is demonstrated in~\cite{pham2018investigating}, where an in-house CSMA/CA MAC control for LoRa is introduced, resulting in an improvement in energy consumption. Overall, achieving a robust and energy-efficient communication link in IoT systems presents challenges that require thorough consideration of various factors, including gateway coverage influenced by channel loss and modulation schemes, channel interference resulting from traffic load, packet size and packet priorities, as well as quality of service and packet latency. In order to tackle these obstacles, the implementation of a custom MAC layer control emerges as a viable solution for IoT designers.

\subsection{Energy Consumption}\label{Sec_Energy}
Energy consumption poses a significant challenge for IoT EDs, particularly in scenarios where power supply cannot be guaranteed for all devices, such as outdoor remote areas. This issue becomes especially critical for early fire detection systems, as their batteries are expected to last for extended periods without frequent replacement. It is essential to make sure that both the sensors and the microcontroller have low energy consumption and that the detection algorithm, whether based on DL/ML, does not consume too much power. To mitigate energy consumption, the sampling frequency of sensors may need to be reduced, and devices may require frequent periods of sleep to conserve power. Alternatively, end devices can offload detection tasks by performing initial threshold detection and transmitting the collected data to a central server for DL or other power-intensive detection algorithms. However, this approach may result in increased energy consumption and packet loss in the communication link. Therefore, striking a balance between the packet size sent to the central node for detection and the resulting latency becomes another challenge in the development of the IoT system. A recent study by~\cite{cryptography6020016} investigated the energy consumption and execution time characteristics of the YOLOv3-tiny algorithm, which is a vision-based fire detection algorithm. The results of the study indicated that the algorithm itself consumes 6.5W (equivalent to 1300mA at 5V) and takes approximately 121 seconds to complete its execution. These findings imply that if an end node relies on a 3000mAh battery, it would only be able to sustain approximately three hours of continuous operation if the YOLOv3-tiny algorithm is employed throughout. Although the operating cycle of the algorithm can be adjusted, it is important to consider that DL methods such as CNNs and RNNs used in these algorithms also require high storage and computing power. Therefore, the idle energy consumption of microcontrollers should not be overlooked. Consequently, IoT systems designers are advised to carefully select sensors, taking into account the energy requirements of detection algorithms, while also ensuring that the nodes have sufficient battery capacity for transmission purposes.

\subsection{Accuracy and Topology}\label{Sec_accuracy}
As mentioned above, to minimise energy consumption in the ED, it is recommended to use energy-efficient detection algorithms. In cases where the decision-making process is based solely on the ED, simpler ML algorithms or statistical threshold detection methods can be employed. However, it is important to note that employing a straightforward detection algorithm, such as threshold detection or simplistic ML classification, can result in potential problems of increased sensitivity and a large number of false alarms. While specific measurement ranges are typically anticipated during unwanted fire incidents, other nonfire-related occurrences can also trigger sensor anomalies. Such anomalies can arise from animals or humans exhaling near the sensors. To improve the precision of fire detection systems, a viable solution is to integrate multiple detection aspects, such as combining vision-based detection and detection based on environmental monitoring sensors, to establish a remote sensing system. Nonetheless, this solution may incur additional energy consumption in each individual end device and additional delay in fire detection. As discussed in Subsection~\ref{Sec_MAC}, the restricted number of EDs resulting from the limited bandwidth of the radio channel requires a throughout design of the system's topology to achieve low latency and reduce packet collisions. In~\cite{abdallah2024centralized} and~\cite{ZHANG20121044}, innovative network topologies are introduced aimed at reducing energy use, enhancing system coverage, and preserving overall system efficiency. These models, while suitable for various IoT networks, assume that IoT gateways can be positioned anywhere in the field without affecting communication quality. In~\cite{9358205}, a more practical approach is suggested, positioning gateways at the forest's perimeter with sensors close to these gateways, showing an improvement in the energy efficiency of the system. However, the delay in detection with this topology model remains unexplored. Therefore, deciding on the optimal number of EDs, vision-based EDs, and the most suitable spacing and angular arrangement between EDs are essential for the IoT wildfire detection system. To address these topological issues, extensive measurement and analysis must be performed. In addition, reducing latency is crucial for prompt detection of wildfires, allowing faster responses when true fires occur. Similarly, it is important to minimise false alarms by evaluating the sensitivity of algorithms, particularly those using edge computing. Higher sensitivity improves a model's ability to detect all true positive instances, which contributes to both reducing false negatives and accelerating response times to actual fires. In~\cite{8753728}, a comparison of the sensitivity of various machine learning models suitable for edge computing is presented. The results show that Naive Bayes achieves the highest sensitivity at 96$\%$, followed by ANN at 95.5$\%$, DT at 93$\%$, and KNN at 93.9$\%$. In contrast, in~\cite{dampage2022forest}, experiments reveal that the sensitivity of employing a statistical threshold is as low as 68.8$\%$.

The challenges involved in implementing an IoT sensor system for early wildfire detection in remote areas can be illustrated by the trade-off between energy consumption, detection accuracy and delay. One of the main challenges is to conserve energy within the system. This can be achieved by using energy-friendly algorithms on the end devices and by sending decisions directly to the main server. However, this approach may not be able to guarantee high accuracy. Furthermore, ensuring detection accuracy requires a comprehensive sensing system that uses multiple aspects for detection and incorporates cross-validation techniques. However, this approach results in increased energy consumption and delay. Alternatively, in the pursuit of avoiding packet collision and achieving rapid response, only the detection results are transmitted to the server, which may lead to compromised accuracy. 
In~\cite{9530253}, researchers performed an analysis of the performance of the IoT network employing both edge computing and cloud computing architectures. In the edge computing example, detection processes are executed on end devices, whereas in the cloud computing framework, these processes are centralised on the main server. The findings reveal that edge computing outperforms cloud computing with an average reduction of 70 ms in detection latency. Furthermore, it is observed that as the packet size increases, the detection latency in cloud computing proportionally escalates. The accuracy of machine learning models ranges between 98.7-99\% for edge computing, and the presence of a large number of sensors introduces an additional processing delay of 2-5 seconds in the system. Although the detection accuracy for cloud computing is not specified, the study also demonstrates that a minimum interval between two consecutive readings is necessary to detect an anomaly. However, this interval may not be achievable with a single sensor reading. In other words, the ML-based anomaly detection model may require multiple successive sensor readings to identify an anomaly, and this problem can be exacerbated if spatial correlation is involved. Overall, there is a complex trade-off between energy consumption, detection accuracy and delay in implementing an IoT sensor system for wildfire detection in remote areas. There is no single solution that is optimal for all situations. The best approach will depend on several factors, including the size of the area to be monitored, the availability of resources and the risk of wildfires.

\section{Suggestions for Future Implementation}\label{Sec_pos_sol}

\begin{figure}
    \centering
    \includegraphics[width=\linewidth]{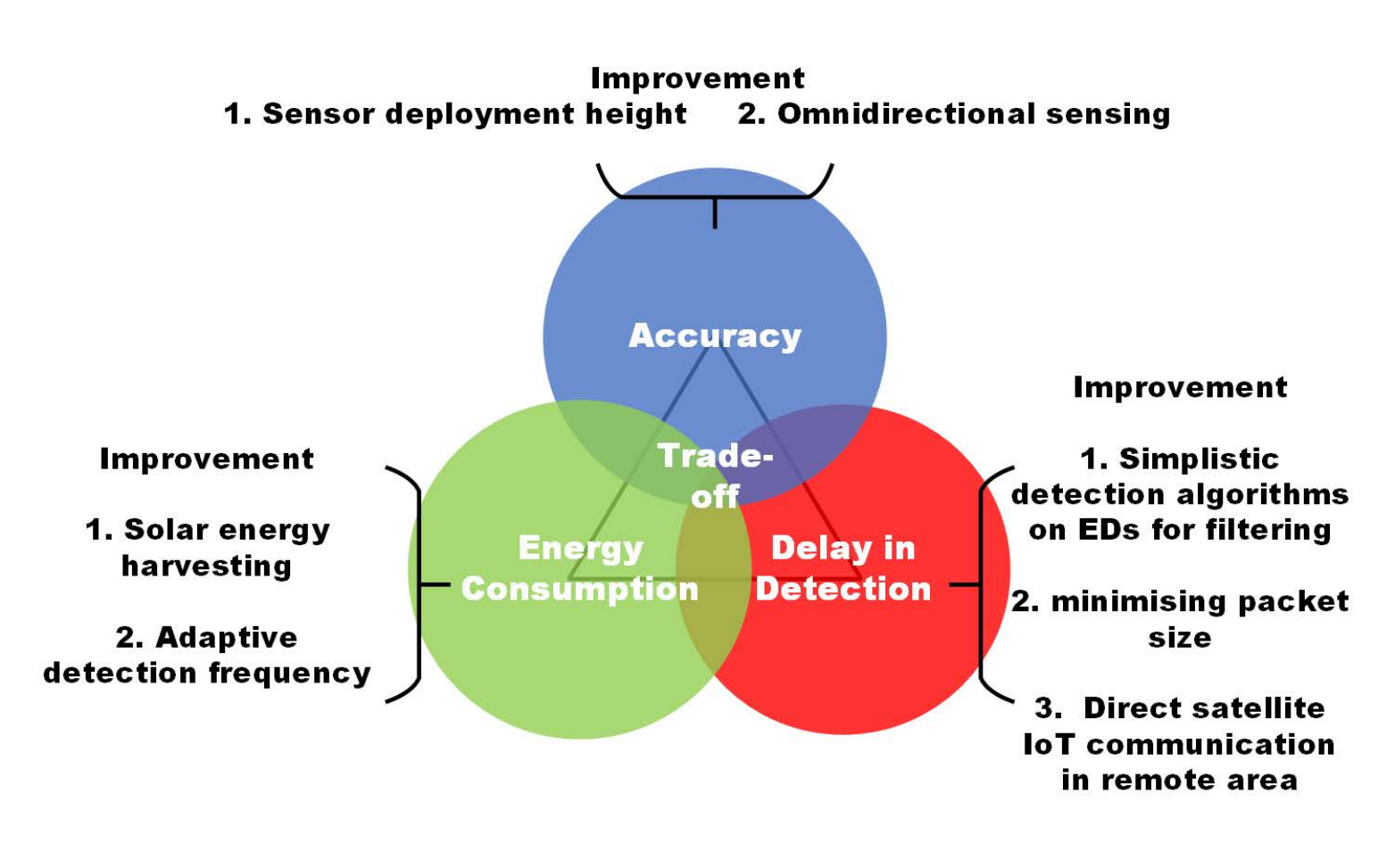}
    \caption{The trade-off of IoT-based wildfire detection systems involves the interrelationship between the accuracy of the sensors, energy consumption, and the delay in detecting wildfires. The accompanying diagram illustrates potential enhancements for each aspect, with the goal of achieving a balance between the overall efficiency and effectiveness of the system in early wildfire detection.}
    \label{fig:tradeoff}\vspace{-1em}
\end{figure}

Despite the challenges encountered, the use of IoT sensing systems can be a valuable tool for early wildfire detection. By integrating detection using environmental data with other imaging detection platforms, such as satellites, UAVs, watchtowers and vision-based IoT, precise early warning systems can be established. In this section, our aim is to provide recommendations for the development of IoT sensing systems addressing issues we discussed in the previous section, with particular emphasis on reducing energy consumption and dealing with the trade-off when selecting sensors and detection algorithms, through the use of a decision-making model. The potential for further development and exploration of IoT ground sensing systems for the early detection of wildfires is also discussed. Fig.~\ref{fig:tradeoff} illustrates the trade-off relationship or the so-called 'dilemma' that arises when implementing the IoT ground sensing system for early wildfire detection. The critical factors to consider for IoT engineers deploying such a system are energy consumption, system accuracy, and detection delay. Suggestions are provided to enhance the capacity of the IoT system in one aspect without significantly compromising other aspects. These suggestions are based on a critical review of the literature combined with the preliminary result of our ongoing research.

\subsection{Addressing Energy Consumption}
\begin{figure}
    \centering
    \includegraphics[width=\linewidth]{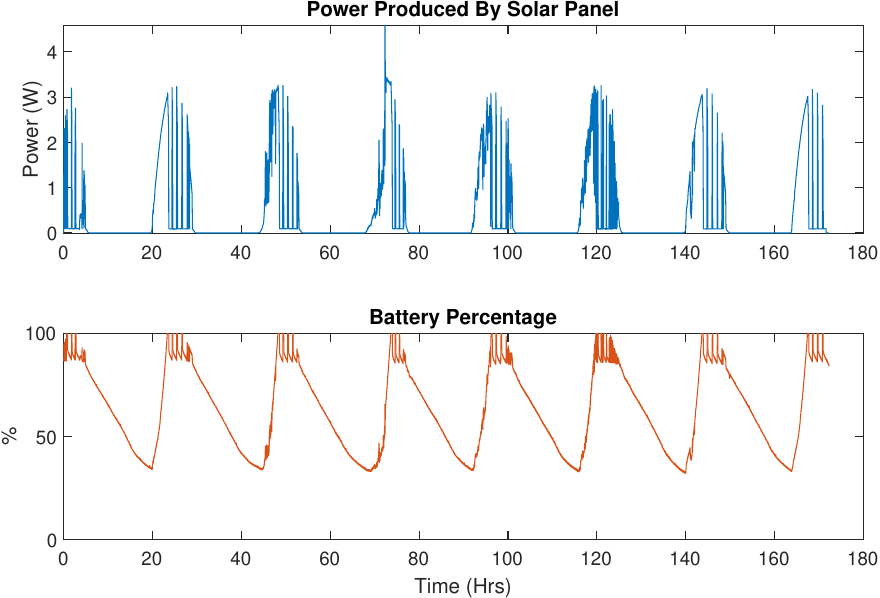}
    \caption{Illustration of the charging dynamics of the end node utilising a polycrystalline solar panel measuring 25x12 cm$^2$ with a capacity of 10 watts, exposed to 1.5 kWh/m$^2$ irradiance, showing daily full battery charging.}
    \label{fig:solardata}\vspace{-1em}
\end{figure}

To reserve energy in the IoT sensing system, it is suggested to use energy harvesting techniques, such as solar energy harvesting, to recharge the ED battery. The battery should last up to the next recharge period. During periods characterised by clear skies, the energy harvested from the solar panel is likely to fully charge the battery. Among all energy harvesting methods, solar energy harvesting stands out as the most practical choice for powering remote IoT systems because of its well-established nature in supplying remote power and its reliability in forested areas, as well as the effective deployment of solar hardware. To investigate the effectiveness of solar power harvesting, we measured the daily energy consumption of the end node and the gateway, selected appropriate solar panel controllers and photovoltaic panels (10 watts for the end node, 30 watts for the gateway) based on an assumed 80\% exposure to the mean daily irradiance, and verified daily energy recharge of the system. Fig.~\ref{fig:solardata} shows the accumulation of energy by the photovoltaic modules of an ED and the battery capacity over a couple of days, thus demonstrating the practicality of incorporating the solar energy harvesting system within the ED framework. To further increase the efficiency of solar energy harvesting, different research projects have suggested and used maximum power point tracking (MPPT) techniques~\cite{electronics9060893, ganesh2013forest}. The MPPT module can auto-track the maximum point of output voltage to optimize the energy conversion. Typically, MPPT can improve energy conversion efficiency by about 10$\%$ to 30$\%$ compared to systems without MPPT under various conditions, in non-ideal circumstances such as partial shading, cloudy weather, or varying temperatures. In a recent study~\cite{LV2024122320}, the radiative cooling film is combined with a solar panel to further increase the efficiency of energy harvesting by adding an energy harvesting feature through radiative cooling to a solar harvesting system and has been proven to outperform the traditional method. However, further investigations are imperative to determine the maximum charging range and to evaluate the efficacy of such applications in outdoor environments.

Beyond energy harvesting, as highlighted in Section~\ref{Sec_lit_review_new}, both MoS and pellistor-type sensors require heating, potentially shortening the lifespan of EDs and increasing maintenance expenses. Therefore, alternative sensing technologies, such as NDIR and FET, which consume only a few microwatts per second, should be taken into account. To further reduce energy consumption in IoT EDs, the sensors can be set to sample at different rates according to weather conditions or the calculated moisture content of fire fuel related to the higher risk of fire on a given day. For example, during extreme heat and aridity conditions, it is advisable to increase the sampling rate to collect a larger number of readings from the devices while reducing it under opposite conditions to conserve the battery. This approach has started to be adopted in recent works such as~\cite{benzekri2020early}. 
\subsection{Improving Sensing Accuracy}
While the effectiveness of the IoT wildfire detection system is highly dependent on the complexity of its detection algorithm, leading to a decision on the selection of detection algorithms, the accuracy is also affected by the deployment approaches used for the sensor nodes. By strategically placing sensors at specific heights and aligning them correctly, it is possible to optimise the detection range, leading to increased sensitivity and ultimately reducing the rate of missed detections. In terms of height and orientation of IoT ED deployment, based on our experiments and the insights from Dryad study~\cite{Dryad} suggest that the end devices should be mounted on trees at a height of 2-3 metres to optimise the detection of TVOC and CO$_2$ in the event of a wildfire. Furthermore, the elevated placement of EDs also improves the LoS between EDs and the gateway. Our experiments have shown that MoS sensors can detect a small fire burning in a 0.7 square metre fire pit within a few minutes from a distance of 50 metres when placed downwind. This implies that the inter-node distances for a grid topology of EDs can be $70.7$ metres. The unpredictability of wildfire ignition points in forested areas makes it difficult to determine the best orientation for individual sensors to maximise their detection capabilities. These sensors are similar to directional antennas, capable of sensing only in one direction. Our experiments have shown that when the sensor is not pointed towards the fire source, the detected gas concentration can be reduced to only 10$\%$ of its original value when facing the ignition point. To address this issue, \cite{10.1145/1134680.1134685} proposed a method that involves placing the sensors in a partially enclosed cage, with the opening facing the ground. This design effectively turns a unidirectional sensor into an omnidirectional one. Our tests confirm that this configuration enables the detection of approximately 50$\%$ of the gas concentration that would be perceived at maximum reception when the sensor is aligned with the fire source. Nevertheless, analysis of the system topology is also crucial. Specifically, adjustments may be necessary when implementing an IoT system in certain forested regions where the presence of fire obstructs the LoS between sensors and it is critical to determine if the system's performance varies from that in open outdoor spaces, taking into account factors like the best sensor elevation and configuration. The effective detection of wildfires is based on the use of multiple sensors, each capable of detecting different fire indicators such as heat, TVOC and CO$_2$. Additionally, to improve the accuracy of detection systems, an effective approach involves incorporating various detection dimensions. This includes expanding the types of input data for decision-making, such as including temperature, RH, and air pressure in the detection model, given that certain activities by humans or animals can cause an increase in RH levels, but in a study by~\cite{s23010169}, the authors summarise the correlation analysis between various types of sensor data, revealing that TVOC, CO$_2$, and PM exhibit limited correlation with fire alarms. Consequently, more research effort is needed to determine the optimal combination of sensor inputs that can effectively enhance the accuracy of ML/DL detection algorithms.

\subsection{Reducing Detection Delay}
The strategies recommended in the preceding section not only enhance detection accuracy but also aid in minimising detection latency by improving sensor sensitivity. However, further measures are needed to mitigate delays resulting from packet loss between EDs and the gateway, and subsequently to the cloud server. To address the challenge of detection delay and energy consumption resulting from packet loss, it is recommended to reduce packet size and transmission frequency. Instead of sending all sensors data back to the server for investigation, research has shown that statistical threshold detection and basic ML algorithms such as KNN or linear SVM can be used for preliminary filtering~\cite{9530090,9332990,dampage2022forest,7794093,forehead2020traffic}. For cloud-based detection, when EDs identify a sequence of alerts related to meteorological and gas parameters, they should activate an alarm and inform fire departments. Subsequently, EDs will transmit their observational data to a central server, where more sophisticated DL techniques, such as RNN or LSTM networks, can be employed for comprehensive cross-verification. The central server can apply spatial correlation strategies to accurately identify the possibility of a fire hazard~\cite{9530253}. Once the fire incident is confirmed, camera modules on site can be used to acquire visual evidence for further examination. In addition, direct satellite communication is viable for remote areas that lack network connectivity. Although it increases gateway energy consumed, costs can be reduced compared to packet relaying. To enhance the dependability of the system and reduce false positives, the generated alert can also be confirmed by cross-validating with watchtowers, UAVs, or satellite entities. In conclusion, the integration and centralisation of IoT technology, UAVs, watchtowers and satellite systems should be coordinated and monitored by fire control centres.

\subsection{Trade-off and Decision Making}

In this section, we have mentioned a range of strategies to enhance the performance of an IoT ground sensing system, focusing on prolonging the lifespan of EDs, optimising detection capabilities and minimising transmission delays. Despite progress in the IoT sensing system, there are still fundamental restrictions due to hardware performance limitations in EDs and energy consumption. This ongoing challenge requires balancing the trade-off between energy consumption, detection accuracy and detection latency. Hence, creating a metric for evaluating these aspects and determining the appropriateness levels can assist IoT engineers in making well-informed decisions. To this end, the multi-criteria decision analysis (MCDA) model is an appropriate decision making framework that can be used to weigh the defined criteria. In a basic MCDA model, assigning a weight to each evaluation criterion enables the calculation of the overall effectiveness, as the weighted sum of all criteria. In this case, the contribution of each criterion is directly proportional to its assigned weight. Additionally, some criteria, such as detection latency or sensor recovery time, can be given an exponential weight, thus introducing a combination of linear and exponential weights into the analysis. This modified MCDA cost $C$ can be expressed by the following equation,
\begin{equation}
    \centering
    C = \sum_i^m W_i x_i + \sum_j^n A_j e^{y_j},
    \label{eq:MCDAform1}
\end{equation}
where the values of the criteria $x_i$ represents the criteria contributing linearly to $C$ and $y_j$ represents the criteria contributing exponentially to $C$. $W_i$ and $A_j$ are the corresponding linear and exponential weights, respectively. The indices $i$ and $j$ range from 1 to $m$ and 1 to $n$ respectively, with $m+n$ being the total number of evaluation criteria. This MCDA framework can be customised to assist in the selection of a suitable sensor based on factors such as cost, energy consumption, response time, and operational environmental ranges.

Table~\ref{tab:MCDA} presents an illustration of how MCDA can be used to evaluate and compare different gas/smoke sensors. In this particular example, we assume that energy consumption, sensor cost, and sensor size have a linear weighted impact on overall cost. Specifically, we assume that an increase of 50 mW in power consumption, an additional \$20 in sensor cost, and an increase of 5 $\mathrm{cm}^3$ in sensor size are deemed to cause a similar level of challenge in hardware implementation or design. Consequently, these factors are assigned costs of 0.02, 0.05, and 0.2, respectively. As minimizing detection delays is of utmost importance, we assign it an exponential weighting. We consider the sensor to be satisfactory if it can detect smoke as changes in CO$_2$, TVOC, or particulates in 90 seconds, with exponential weight $A_1$ equal to $\frac{1}{e^{90}}$. The SGP30 sensor has the lowest weighted sum cost among the sensors, making it the most suitable candidate for this particular MCDA.
\begin{table}
    \centering
        \caption{Illustration of using mcda in evaluating the performance of gas/smoke sensors}
    \resizebox{\linewidth}{!}{\begin{tabular}{l l l l l l}
    \toprule
         \multirow{3}{*}{\textbf{Sensor}} & \textbf{Criterion $x_1$} & \textbf{Criterion $x_2$}& \textbf{Criterion $x_3$}& \textbf{Criterion $y_1$}& \multirow{3}{*}{\textbf{Weighted sum}}\\
         & $W_1 = 0.02$ & $W_2 = 0.05$ & $W_3 = 0.2$ & $A_1 = \frac{1}{e^{90}}$\\
         & Energy consumption [mW] & Price [\$AUD] & Size [cm$^3$]& Response time [sec]  &   \\
         \midrule
        MQ2~\cite{MQ2} & 343 &  12 & 19& 90 & 12.26\\
        PMS5003~\cite{pms5003} & 550 & 78& 40& 30 & 22.9\\
        SCD40~\cite{scd40} & 50.2  & 130 & 4.032& 80 & 8.31\\
        SGP30~\cite{SGP30}  & 150   & 35 & 1.083& 30 & 4.97\\
        BME680~\cite{BME688} & 49.8   & 30 & 0.8424& 110 & 4.85$\times 10^8$\\
         \bottomrule
    \end{tabular}}
    \label{tab:MCDA}\vspace{-1em}
\end{table}

Additionally, the MCDA can be used to determine the most effective anomaly detection method, whether point-wise, time-series, or ML/DL detection techniques. As unsupervised detection algorithms cannot be evaluated using the F1 scoring system as in supervised learning, the MCDA can also be used to effectively compare the two methods. This can be done by taking into account criteria such as the delay from the ignition point until detection and the number of false alarms. Last but not least, the framework can be used to evaluate the effectiveness of on-board processing in EDs or processing on the centralised server.

\section{Conclusion}\label{Sec_Conc}
An overview of the state-of-the-art of IoT ground sensing wildfire detection systems is presented in this article. The underlying principles of various vision-based detection algorithms, as well as the working principles of multiple environmental sensing units and various detection algorithms employed in ground sensing anomaly detection, are explained. A detailed discussion of the challenges encountered during the development of an IoT ground sensing system for wildfire detection is also provided. Based on the findings of the literature review and our own experimental observations, suggestions for the development of a robust and reliable IoT early wildfire detection system are proposed, along with suggestions for future research possibilities.
\bibliographystyle{IEEEtran}
\bibliography{references}
\end{document}